\documentclass[11pt]{article}

\usepackage[]{acl}

\usepackage{times}
\usepackage{newtxmath}
\usepackage{latexsym}
\usepackage{url}
\usepackage[T1]{fontenc}

\usepackage[utf8]{inputenc}

\usepackage{microtype}

\usepackage{inconsolata}
\usepackage{booktabs}
\usepackage{amsmath}
\usepackage{xcolor}
\usepackage{array}
\usepackage{longtable}

\usepackage[table]{xcolor}
\usepackage{fontawesome5}
\newcommand{\ghicon}{%
  \raisebox{-0.12em}{\scalebox{1.2}{\faGithub}}%
}

\newcommand{\ghlink}[2]{
\href{#1}{
\ghicon\hspace{0.35em}\texttt{#2}}
}

\newenvironment{itemize*}
  {\begin{itemize}
    \setlength{\itemsep}{0.9pt}
    \setlength{\parskip}{0.9pt}
    \setlength{\topsep}{0.9pt}}
  {\end{itemize}}

\usepackage[most]{tcolorbox}

\newtcolorbox{promptbox}{
    colback=gray!8,
    colframe=gray!50,
    boxrule=0.5pt,
    arc=2mm,
    left=2mm,
    right=2mm,
    top=1mm,
    bottom=1mm
}

\usepackage[nameinlink]{cleveref}

\crefformat{section}{\S#2#1#3} 
\crefformat{subsection}{\S#2#1#3}
\crefformat{subsubsection}{\S#2#1#3}

\newcommand{\tracehl}[1]{%
  \colorbox{green!20}{#1}%
}
\usepackage{multirow}
\usepackage{graphicx}
\usepackage{todonotes}
\makeatletter
\newcommand*\iftodonotes{\if@todonotes@disabled\expandafter\@secondoftwo\else\expandafter\@firstoftwo\fi} 
\makeatother

\newcommand{\projName}{InfiniSearch} 
\newcommand{\infinigram}{\textit{infini-gram}}

\title{\projName: \\ Retrieval from LLM-Scale Corpora Beyond Lexical Matching}

\author{Mattes Ruckdeschel \quad Toine Bogers \quad Anna Rogers \\
  IT University of Copenhagen \\ 
  \texttt{mruc@itu.dk} \quad \texttt{tobo@itu.dk} \quad \texttt{arog@itu.dk} \\
  \ghlink{https://github.com/MattesR/InfiniSearch}{MattesR/InfiniSearch}
  }

\begin{document}

\newpage
\setcounter{page}{1}
\maketitle

\begin{abstract}

To better understand and evaluate modern Large Language Models (LLMs), we need to be able to explore their training data. A key problem is that 
traditional retrievers do not scale to LLM dataset sizes. Byte Suffix Arrays (BSAs) offer fast retrieval that scales, but suffer from strict lexical n-gram matching. We introduce \projName{}, a BSA-based first stage retriever that addresses the problem of lexical matching with keyword expansion, while preserving the scalability of BSAs. We validate its performance in a three-staged retriever setup on a range of information retrieval benchmarks, and we show its potential to support research on LLM-scale corpora in a qualitative study on a 4.6T token corpus.
\end{abstract}

\section{Introduction}

The success of modern LLMs comes down to the paradigm of scaling, both in terms of parameter counts and training data \cite{KaplanMcCandlishEtAl_2020_Scaling_Laws_for_Neural_Language_Models}. 
At this scale, data management alone becomes a challenge, including data filtering \cite{liDataCompLMSearchNext2025} and provenance \cite{longpreBridgingDataProvenance2024}.
Data-dependent research on memorization  \cite{wangGeneralizationVsMemorization2024} and data attribution \citep{liuOLMoTraceTracingLanguage2025} also becomes prohibitively costly. %
There are many proposals for how to \textit{document} datasets better \cite[][\textit{i.a.}]{mitchellModelCardsModel2019,maroneDataPortraitsRecording2023,jerniteDataGovernanceAge2022,longprePretrainersGuideTraining2024}, as well as initiatives to document some publicly available resources \cite[][\textit{i.a.}]{gaoPile800GBDataset2020, penedoFineWebDatasetsDecanting2024, soldainiDolmaOpenCorpus2024,lambertTulu3Pushing2025}. However, there are currently no scalable frameworks that can go beyond descriptive statistics and lexical search. %

Suffix-arrays \cite{manberSuffixArraysNew1993} are a fast, scalable solution for exact-substring matching \cite{leeDeduplicatingTrainingData2022}. Recently, \citet{liuInfinigramScalingUnbounded2024} built \infinigram{}, an engine based on byte-suffix-arrays (BSAs). \textit{Infini-gram} supports search and retrieval over tokenized pre-training datasets, returning counts for unbounded n-grams within milliseconds. This makes \infinigram{} a promising candidate for scaled retrieval, but to date this direction has not been explored. A key challenge is that BSA matches are binary: if a document does not contain the search term, it is not retrievable. It is thus imperative to find a suitable generation scheme for index-queries with sufficient precision and recall for arbitrary queries.

Our contributions are as follows. 
We present \projName{}, a first-stage retriever based on \infinigram{}, which overcomes the dependency on verbatim matching through LLM-based keyword expansion and a data-driven query generation scheme. We show that our solution yields competitive results on common retrieval benchmarks to solutions with similar hardware footprint. We also conduct a qualitative analysis of retrieval on a 4.6 trillion token dataset, which illustrates how \projName{} can support in-depth analysis of LLM corpora, prohibitively large to study with current methods. Our implementation and the code to reproduce all reported experiments, including keyword expansions and qualitative analysis data, publicly available.

\section{Related Work}
\subsection{LLMs for Query Expansion}
Keyword expansion is a common solution to the verbatim matching problem \cite{carpinetoSurveyAutomaticQuery2012}.
There is recent work on query expansion with LLMs. It relies on the assumption that the models likely learned \textit{some} information about \textit{common} queries during their extensive pre-training \cite{zhuLargeLanguageModels2025}. This line of work includes:
\begin{itemize*}
\item 
\citet{jagermanQueryExpansionPrompting2023} extends the method of Pseudo-Relevance-Feedback\footnote{See Chapter 9 of \citet{schutze2008introduction} for an overview of the method from an NLP perspective.} \cite{robertsonTERMSELECTIONQUERY1990a}, which retrieves \textit{pseudo-relevant} documents (i.e. documents assumed to be relevant based on their retrieval rank, without verified relevance judgments) and mines them for additional relevant terms. \citet{jagermanQueryExpansionPrompting2023} replace the retrieval of pseudo-relevant documents by generating them. 
\item \citet{gaoPreciseZeroShotDense2023} do something similar for dense retrievers: first, queries are answered by an LLM, and these \textit{hypothetical} documents are used as queries for a dense retriever. 
\item Query2Doc \cite{wangQuery2docQueryExpansion2023} simply extends the query for either dense or sparse retrieval with a pseudo-document from an LLM.
\item Word2Passage \cite{choiWord2PassageWordlevelImportance2025} uses LLMs to generate expansions in a three-way setup: they expand a query at the word-, sentence-, and passage-level, and score the importance of words using corpus characteristics, to weigh the terms of the query expansion. 
\end{itemize*}
Our work introduces keyword expansion for the particular challenges of searching in BSAs, where expansions
have to be atomic, so as to increase the likelihood of matches in the index. Query expansion presents a trade-off \cite{wellerWhenGenerativeQuery2024a}: while it helps to enhance recall (especially for weaker retrievers), it also introduces noise, which makes it hard to efficiently rank truly relevant results. We address this trade-off by analyzing different query construction strategies to find query types with high precision.
SEAL \cite{bevilacquaAutoregressiveSearchEngines2022} uses an FM-Index \cite{ferraginaOpportunisticDataStructures2000}, which is a compressed BSA and an autoregressive LM to generate n-gram identifier for retrieval on a small dataset (13.4 GB). This approach, similarly to ours, combines an LM with a scalable retriever, but it depends on fine-tuning a model with gold data. Such data is not available for pre-training corpora, and the method would not scale to it.

\begin{figure*}[ht!]
    \centering
    \includegraphics[width=\linewidth]{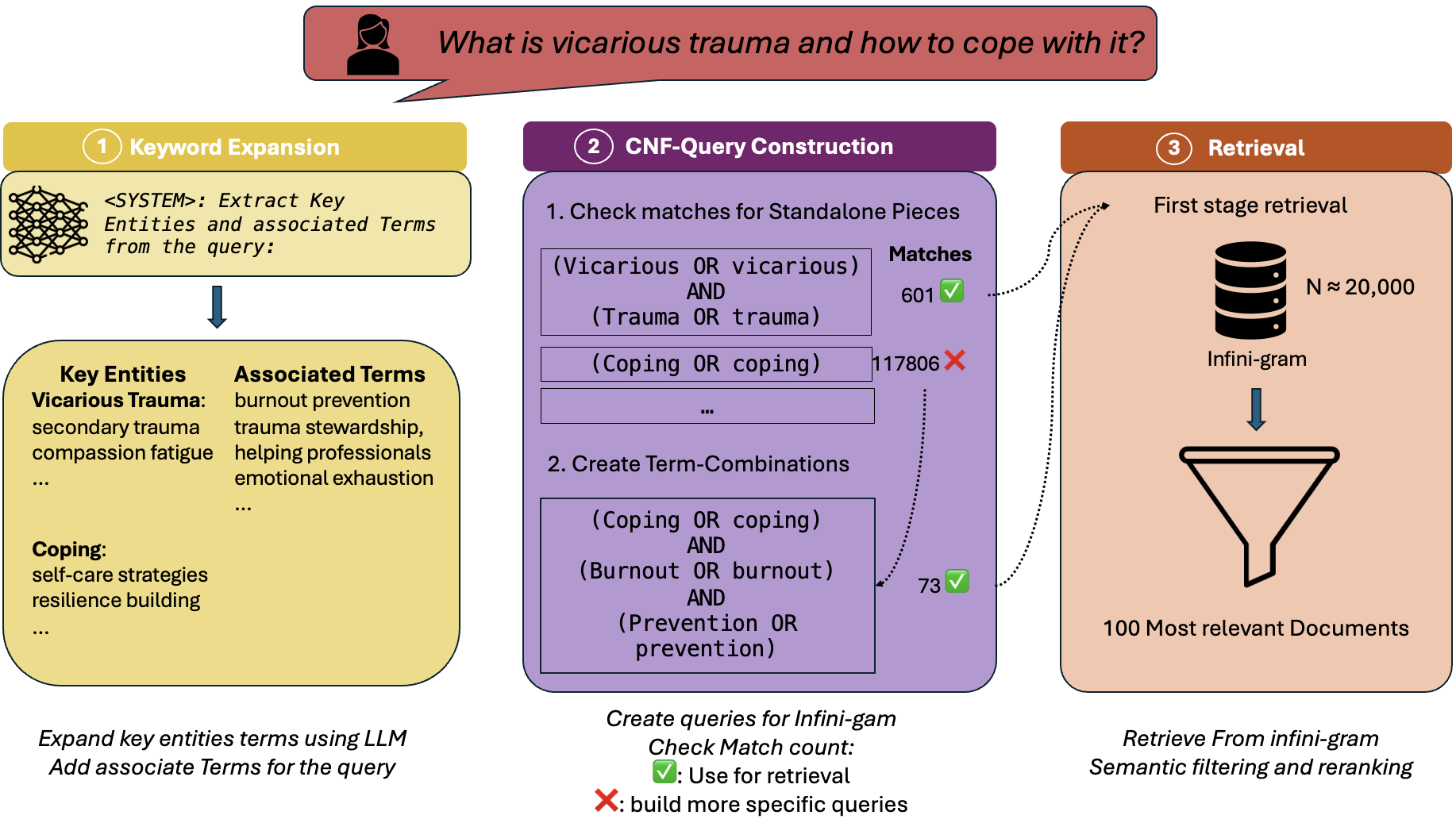}
    \caption{The \projName{} pipeline: LLM extracts key entities from a query and suggests associated terms (\autoref{sec:kw_exp}). From these, CNF queries for a Byte-Suffix-Array, \infinigram{}, are constructed (\autoref{sec:cnf_quer}). Queries that are specific enough are used for first-stage retrieval from the Byte-Suffix Array and the results are filtered in two additional stages (\autoref{sec:retr}).}
    \label{fig:figure_1}
\end{figure*}

\subsection{Indices for pre-training data}
There are few publicly available LLM-scale indices, or recipes for building such indices. 

The \textit{ROOTs search tool} \cite{piktusROOTSSearchTool2023} targeted the 1.6Tb pre-training data of BLOOM model \cite{workshopBLOOM176BParameterOpenAccess2023}. It combines a Lucene index with a BSA for fuzzy matching of the training data. %
\textit{`What's in my big data'} project \cite{elazarWhatsMyBig2023} relied on a Lucene-based index for investigating various large corpora. It only enabled exact matching. Both studies explicitly mention performance issues, and %
these indices are either no longer publicly available or have to be self-hosted. It is unclear to what extent their performance issues were due to considerations of hardware and serving costs.

To the best of our knowledge, the \infinigram{} API is currently the only publicly available and searchable index of an LLM-scale pre-training corpus. It is an efficient implementation of a BSA, which comes with a search engine API. Our qualitative assessment of the retrieval capabilities of \projName{} is based on a 22.4 TB index, which well surpasses the indices used in previous work.

\section{Engine Description}
Figure \ref{fig:figure_1} gives an overview over the entire retrieval pipeline. In this section we describe its key components: keyword expansion (\S\ref{sec:kw_exp}), CNF-query construction (\S\ref{sec:cnf_quer}), and retrieval (\S\ref{sec:retr}). 

Our pipeline was built with the MSMarco v2.1 corpus \cite{pradeepRagnarokReusableRAG2024} as the main development resource (specifically on the queries from the 2024 TREC Rag challenge \cite{thakurSupportEvaluationTREC2025}). The evaluation on this corpus is described in \cref{sec:trec_results}. We then report results on BEIR \cite{thakurBEIRHeterogenousBenchmark2021} benchmarks of three other types, which we did not use in development (\cref{sec:beir_results}).

The underlying index is \infinigram{} \cite{liuInfinigramScalingUnbounded2024}, a byte-suffix-array index built on llama-2 tokens \cite{touvronLlama2Open2023}. Metadata and full text of documents can also be stored, using lookup files that store the offset of documents with respect to the BSA.
We did not conduct any preprocessing before indexing.%

\subsection{Keyword expansion}
\label{sec:kw_exp}

The key objective in keyword expansion is to go beyond verbatim matching. In many instances retrieval could be improved by considering synonyms and associated terms. In the example shown in \autoref{fig:figure_1}, \textit{vicarious trauma} has synonyms \textit{secondary traumatic stress} and
\textit{compassion fatigue}\footnote{\url{https://en.wikipedia.org/wiki/Vicarious_traumatization}}, as well as associated terms such as \textit{burnout prevention}.
We rely on LLMs for identifying such semantic expansions for our underlying BSA. The LLM is prompted to identify and expand the key entities from the query, i.e. the noun phrases that are central to the information need. %
Our prompts aim to expand key entities with five terms, and find an additional ten associated terms per query, ordered by their importance to the query. Pilot experiments showed that these settings suffice for our purposes, but we expect the optimal numbers to differ for specific applications. The full prompt is provided in \cref{ap:prompt}.

In scope of this work, we tested four proprietary and open-weights models for effectiveness in keyword expansion: Claude-Opus 4.6 \cite{ClaudeOpus46},  Claude-Sonnet 4.6 \cite{IntroducingSonnet46},
Llama-3:70B \cite{grattafiori2024llama3herdmodels} and Gemma-4:32B \cite{Gemma4Model}.
Experiments with open-weight models were conducted 
using openrouter\footnote{\url{openrouter.ai}}, and with the others through the official APIs.
We evaluate them on the maximum recall that can be achieved using the generated keywords. We additionally report the Conditional Value at Risk (CVaR), which is the mean recall for the worst 10\% of queries, to check the robustness of the generated keywords.
\autoref{tab:max_recall_llm_compare} shows the results for finding relevant documents for the topics from the test set of TREC 2024 RAG track \cite{thakurSupportEvaluationTREC2025}. 
Claude-Opus4.6 was the best performing model, achieving an average recall of 85.0\% over all 301 queries of TREC 2024 RAG. The mean of the worst 10\% of queries is 0.454. Claude-Opus 4.6 performed significantly better than the other (smaller) models. Gemma-4 performs as well as Sonnet, and Llama 3:70B comes the closest to the performance of Opus in general recall. However, Opus has a significantly higher CVaR, and a lower standard deviation, implying that its performance is more robust with respect to the query. 

\begin{table*}[t]
\footnotesize
\centering
\begin{tabular}{@{}ccccccc@{}}
\toprule
                    & \multicolumn{2}{c}{\textbf{Ablations with Opus 4.6}}                        & \multicolumn{4}{c}{\textbf{Recall with keyword expansions and associated terms}}                    \\ \midrule
\textbf{Metric}     & Opus$_\mathit{kws}$ & \multicolumn{1}{c|}{Opus$_\mathit{eo}$} & Opus 4.6       & Sonnet 4.6 & Llama 3:70B & Gemma4:32B \\ \midrule
Mean Recall         & 0.696               & \multicolumn{1}{c|}{0.823}              & \textbf{0.850} & 0.761      & 0.805       & 0.761f     \\
Std Recall          & 0.259               & \multicolumn{1}{c|}{0.192}              & \textbf{0.167} & 0.230      & 0.203       & 0.225      \\
Median Recall       & 0.772               & \multicolumn{1}{c|}{0.898}              & \textbf{0.91}  & 0.841      & 0.868       & 0.833      \\
CVaR                & 0.159               & \multicolumn{1}{c|}{0.378}              & \textbf{0.454} & 0.256      & 0.321       & 0.266      \\
Documents Retrieved & 2,103,468           & \multicolumn{1}{c|}{3,719,155}          & 3,774,197      & 2,192,200  & 2,410,071   & 2,268,742  \\ \bottomrule
\end{tabular}
\caption{Recall results on the test set of TREC RAG 2024, with keyword expansions and associated terms by Opus 4.6, Sonnet 4.6, Llama 4:70B and Gemma 4:32B. For the highest-performing model Opus4.6, we also provide ablations: Opus$_\mathit{kws}$ shows the recall of the keywords without expansions and associated terms, and Opus$_\mathit{eo}$ ablates the recall with only keyword expansions, and no associated terms.}
\label{tab:max_recall_llm_compare}
\end{table*}

We also conducted an ablation on
the effectiveness of both the keyword expansions and the addition of associated terms. We use the key entities from Opus but create only queries from the key entities (Opus$_{kws}$), or omit associated terms from query creation (Opus$_{eo}$).
Without any expansions (Opus$_{kws}$), the recall is 69.6\%, a value that all LLM expansions surpass. The recall increase for Opus when adding the associated terms is only 2.7\%, while at the same time adding 7.6\% to CVaR. This indicates that for many queries, the keyword expansions already ensure most of the achieved recall, but the associated terms add robustness in cases where the keyword expansions are insufficient.

For all LLMs, we can see that the extended terms alone are not specific enough: we retrieve millions of documents from the index to find 228 relevant documents on average. Therefore, we investigate next how best to construct CNF queries to retrieve documents from the index in a more specific way.

\begin{figure}[!t]
    \centering
    \includegraphics[width=0.8\linewidth]{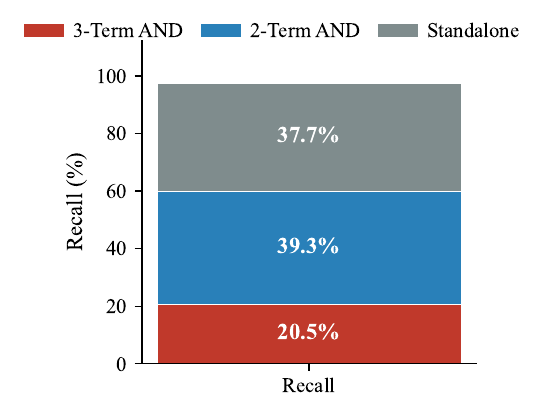}
    \caption{Contribution of standalone, 2-term and 3-term queries to the overall recall. 37.7\% of documents can only be found by a standalone keyword search. These types of searches make the pipeline generally too noisy.}
    \label{fig:query_analysis}
\end{figure}

\subsection{CNF Query Construction}
\label{sec:cnf_quer}

The search engine of \infinigram{} allows for two types of queries: exact matches, and \textit{CNF queries}, which consist of logical expressions in conjunctive normal form. CNF queries can be constructed by conjunctions ({\tt AND}) of clauses connected by disjunctions ({\tt OR}), as seen in Figure \ref{fig:figure_1}. The maximum token distance between these clauses can be specified. 

We use CNF queries to overcome two main weaknesses of the underlying index. Given an entity, such as \textit{Vicarious Trauma}, we can reduce sensitivity to capitalization ({\tt vicarious OR Vicarious}). We also loosen the requirement of exactly matching multi-word entities to create a \textit{CNF piece}: {\tt (Vicarious OR vicarious) AND (Trauma OR trauma)}, with a maximum distance of 10 tokens between parts of the entity.

An effective first-stage retriever should only retrieve as many documents as can be processed by the second-stage retrieval component in a reasonable time. By combining terms from different key term expansions, we can build more specific CNF queries.
While queries to the infinigram index are fast, their latency rises with the number of elements in the CNF query. Hence, there is a tradeoff between the specificity and the latency of a query. Additionally, it is infeasible to try all possible combinations of queries from the expansion.\footnote{Estimates for the number of queries for different numbers of extracted keywords are provided in \cref{ap:ap_queries}} Therefore, we only investigate multi-term queries that combine two or three terms. 

\begin{figure}[!t]
    \centering
     \includegraphics[width=\columnwidth]{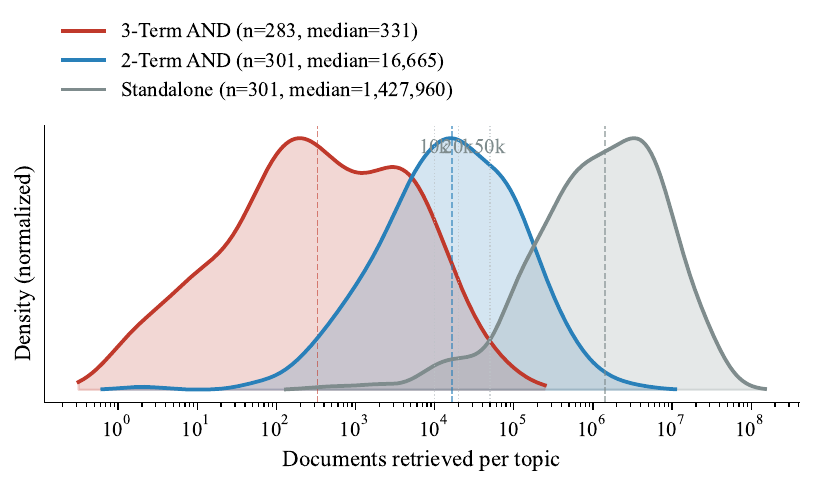}
     \caption{Retrieval distribution for 3-Term-AND, 2-Term-AND, and standalone queries, on a log scale.}
    \label{fig:engine_count}
 \end{figure}

Figure \ref{fig:query_analysis} shows the maximum recall possible with one/two/three-term CNF queries among all relevant documents that match at least one expanded term. We can see that 37.7\% of retrievable documents only contain a single term and would not be found by any combined query. From \autoref{fig:engine_count}, we see that all types of queries have a wide spread of documents that they potentially retrieve. A keyword like \textit{career} has millions of matches, while \textit{vicarious trauma} has only 600. Retrieval from the BSA must therefore be able to handle both extremely broad and extremely specific queries.

When generating CNF queries for retrieval, we count the matches of the individual parts of a query and build queries depending on the matches. For our setup, we found empirically that 2-term-AND queries are the most effective, since there are fewer to build, and they cover all documents identifiable with 3-term-AND queries, albeit with more noise.

Following these considerations, we build the CNF queries for document retrieval in three stages. In each stage, we build increasingly specific CNF queries, count their matches against the index, and add CNF queries below the threshold of document budget $B$ to the pool of queries for retrieval.

\textbf{Stage 1.} We first count the the document matches for all standalone queries, and add standalone terms that have below $max_{sa}$ matches in the index directly for retrieval. This eliminates them from CNF-query construction in the next stage, saving on the number of queries. 

\textbf{Stage 2.} We build 2-Term-AND queries from keyword expansions. \autoref{tab:max_recall_llm_compare} shows that the key term expansions are more important for high recall than the associated terms. We only combine associated terms in 2-Term-AND queries to the first keyword, assumed to be the most discriminative. This further limits the number of possible queries.

\textbf{Stage 3.} We check for $B$. If the results of second stage match more than $B$ documents, we sort the queries by their number of matches, and add queries until $B$ is reached. We only add queries constructed with associated terms after adding the direct key terms and their expansions. If less than $B$ documents are in the pool, we fill the pool with the most specific
queries, beginning with queries containing key entities directly and following with queries that contain their expansions.
 
In practice, the same document often matches multiple queries, and hence the match count reflects only an upper bound of the number of retrieved documents. We set $B$ to 40,000 matches, which retrieved 25,662 documents per query on average.

\subsection{Retrieval and reranking}
\label{sec:retr}
Once the pool of CNF queries is constructed, we find the corresponding documents from \infinigram{}. First we find corresponding metadata, and then we decode the stored document tokens back to text for processing in subsequent steps.

The novelty of our retrieval pipeline lies in the use of BSAs as the underlying index for first-stage retrieval, which makes retrieval from LLM-scale corpora possible. We subsequently use off-the-shelf models for a semantic filtering and reranking. Our results are obtained with a small sentence transformer \cite{reimersSentenceBERTSentenceEmbeddings2019}, \textit{all-MiniLM-L6-v2} for filtering down to the 1,000 most similar documents, and a cross-encoder \textit{ms-marco-MiniLM-L-12-v2} for last reranking step, which orders the top 100 results from the previous stage by their relevance to the query.\footnote{We used the models provided at \url{sbert.net}.} 

We deliberately opted for relatively small models for retrieval and reranking, since they add minimal latency and hardware requirements, and allow for fairer comparisons against baselines in our experiments. Additionally, this pipeline can be served without GPU access. Our results should hence be interpreted as a lower bound: when resources are not a concern, it is of course possible to use more high-performing models for filtering and reranking.

\section{Results}
\subsection{TREC RAG 2024}
\label{sec:trec_results}

\begin{table}[!t]
\centering
\footnotesize
\setlength{\tabcolsep}{4pt}
\begin{tabular}{@{}ccccc@{}}
\toprule

        & \multicolumn{4}{c}{Retriever}            \\ \midrule
Metric  & BM25 & BM25$_\mathit{exp}$ & \projName{} & Webis-01 \\ \midrule
NDCG@10 & 0.47 & 0.52     & 0.54        & \textbf{0.66}    \\
R.@100  & 0.14 & 0.12     & 0.22   & -        \\
R@1000  & 0.35 & 0.38     & 0.39        & \textbf{0.81}     \\ \bottomrule
\end{tabular}
\caption{Retriever metrics on the test set of the TREC RAG 2024 test set. BM25$_\mathit{exp}$ uses the generated keywords for query expansion.
Webis-01 results are for the best run reported by \citet{Gienapp2024WebisAT}.}
\label{tab:trec_results}
\end{table}

\begin{table*}[]
\centering
\footnotesize
\begin{tabular}{@{}ccccc|cccc@{}}
\toprule
              & \multicolumn{4}{c|}{\textbf{NDCG@10}}                         & \multicolumn{4}{c}{\textbf{Recall@100}}                           \\
Benchmark     & BM25           & SPLADE         & Contriever & InfiniSearch   & BM25           & SPLADE         & Contriever     & InfiniSearch   \\ \midrule
TREC-Covid    & 0.656          & 0.727          & 0.596      & \textbf{0.746} & 0.114          & 0.128          & 0.091          & \textbf{0.130} \\
FEVER         & 0.753          & 0.788          & 0.758      & \textbf{0.838} & 0.931          & 0.946          & 0.949          & \textbf{0.960} \\
Climate-FEVER & 0.213          & 0.230          & 0.237      & \textbf{0.254} & 0.436          & 0.521          & \textbf{0.575} & 0.558          \\
SciFact       & 0.665          & \textbf{0.704} & 0.677      & 0.701          & 0.908          & \textbf{0.935} & 0.947          & 0.894          \\
ArguAna       & 0.414          & \textbf{0.520} & 0.446      & 0.233          & 0.943          & 0.974          & \textbf{0.977} & 0.809          \\
Touché 2020   & \textbf{0.367} & 0.247          & 0.204      & 0.253          & \textbf{0.538} & 0.471          & 0.442          & 0.493          \\
FiQA-2018     & 0.236          & \textbf{0.347} & 0.329      & 0.344          & 0.539          & 0.631          & \textbf{0.656} & 0.581          \\ \bottomrule
\end{tabular}
\caption{NDCG@10 and Recall@100 of baselines compared to InfiniSearch. InfiniSearch does well on fact-retrieval benchmarks, but struggles on very long (SciFact), or very short (FiQA, Touché) inputs.}
\label{tab:comparisons}
\end{table*}

As a sanity check, we compare the performance of \projName{} to BM25 baselines on TREC RAG 2024 test set. We use Pyserini for creating Lucene-indices and for retrieval \cite{Lin_etal_SIGIR2021_Pyserini}.
We use the same parameters as \citet{pradeepRagnarokReusableRAG2024}, but with the same reranker as our \projName{} pipeline. We also include the best reported retrieval results by Webis-01 \cite{Gienapp2024WebisAT}. Their pipeline fuses the results from sparse and dense retrievers, which are then reranked by RankZephyr, a 7 Billion Parameter LLM-based reranker \cite{pradeepRankZephyrEffectiveRobust2023}. We test against the official \textit{query relevance judgments} (QRELs), which were generated using UMbrela \cite{upadhyayUMBRELAUMbrelaOpenSource2024}, a relevance estimation based on GPT-4o.

\autoref{tab:trec_results} shows that \projName{} performs better than BM25 with reranking. BM25 also benefits from using the generated keywords as expansions. The performance reported by \citet{Gienapp2024WebisAT} is significantly higher, especially in recall. 

The results show that \projName{} performs well as a retriever. As expected, it is not on par with sophisticated pipelines like Webis-01, but it serves different goals: a system like Webis-01 would not be usable on the scale of LLM training corpora.

\subsection{BEIR Benchmarks}
\label{sec:beir_results}

To reiterate, the TREC RAG dataset, results of which are reported in \cref{sec:trec_results}, is the main resource we used in the pipeline development. Its results should hence only be interpreted for validation that \projName{} does work as a retriever. We now consider benchmarks unused in development.%

For that, we evaluated \projName{} on various evaluation datasets from the BEIR Information Retrieval suite. Specifically, we chose fact-checking benchmarks \cite{thorneFEVERLargescaleDataset2018, diggelmannCLIMATEFEVERDatasetVerification2021, scifact}, argument retrieval \cite{bondarenkoOverviewTouche20202020, arguana}, and question-answering tasks \cite{fiqa}, to cover a wide range of possible applications, dataset sizes, and query types. We report all evaluations that we conducted.

We compare the results of \projName{} to the reported baselines \cite{kamalloo:2024}: a lexical approach (BM25), SPLADE++ \cite{formalDistillationHardNegative2022}, an enhanced version of SPLADE, and Contriever \cite{izacardUnsupervisedDenseInformation2022}, which is a dense retriever that was trained with contrastive learning from MS Marco examples. 

Neither SPLADE nor Contriever were fine-tuned specifically for these tasks. Likewise, we use \projName{} in the described setup without any changes to the prompt used for keyword expansion, or the hyperparameter settings that we used for the TREC RAG 2024 results. 

Table \ref{tab:comparisons} shows the performance with respect to NDCG@10 and Recall@100.\footnote{See Table \ref{tab:retrieval_results_full} in \ref{sec:retrieval_details} for additional metrics.} NDCG@10 measures the ranking quality and the concentration of relevant documents within the top 10 results, and Recall@100 indicates how good
is the pool of found documents in the top 100. 

Our results suggest that \projName{} performs well on a varied set of standard information retrieval benchmarks without any further information or fine-tuning on the benchmarks' data. This validates the core idea that BSAs combined with LLM-based keyword expansion and a further filtering of results can be used to retrieve relevant documents for a variety of tasks. We expect that the performance would increase with task-specific keyword expansions, and feedback from annotated data.

\section{Efficiency}
\label{sec:efficiency}

We analyze the efficiency of \projName{} compared to BM25, to compare their resource cost regarding both index construction and querying the system. Ideally, the resource requirements would be comparable to BM25 in the smaller IR benchmarks, while remaining scalable to LLM-size corpora. 

For BM25, we build a Lucene index that stores positions and documents so that both systems can return document text for reranking. BEIR measurements were
taken on a laptop (Apple M4 Pro, 24\,GB RAM) with the same thread count (10)
and memory budget (12\,GiB) for both systems; reported times are medians.
MSMarco measurements were taken on a server with a budget of 60 CPU cores and 256GB of RAM.

\subsection{Index Construction}

\begin{table}[t]
\centering
\small
\setlength{\tabcolsep}{4pt}
\begin{tabular}{lrrrr}
\toprule
 & \multicolumn{2}{c}{Index} & \multicolumn{2}{c}{Query latency (s)} \\
\cmidrule(lr){2-3}\cmidrule(lr){4-5}
 & Build (h) & Size (GB) & median & p90 \\
\midrule
InfiniSearch & 1.22 & 326 & 32.7 & 68.2 \\
BM25 & 0.22 & 128 & 0.39 & 0.64 \\
\bottomrule
\end{tabular}
\caption{Efficiency on MS~MARCO v2.1.
First-stage latency on 50 TREC RAG 2024 topics with precomputed keyword
expansions, including reading the documents. BM25: Lucene index with
positions and stored documents. BEIR results can be found in
Appendix~\ref{ap:ap_efficiency}.}
\label{tab:efficiency}
\end{table}

\autoref{tab:efficiency} compares the efficiency of InfiniSearch and a Lucene-Index for BM25 on MS~MARCO v2.1 , infini-gram
takes 1.2\,h against 0.2\,h for Lucene. Three
quarters of this is tokenization. Tokenization time can realistically be cut by using a fast tokenizer and more parallelization.
Building the suffix arrays takes 18\,min,
comparable to Lucene's 13\,min. The index is $2.5\times$ larger (326 vs.\
128\,GB), mostly due to the suffix arrays (165\,GB) and per-segment metadata
(94\,GB), which is not compressed in \infinigram.
For the small BEIR indices, construction is comparable.
On Touch\'e-2020 (382k documents, 160M tokens), infini-gram takes
49\,s against 35\,s for Lucene. 81\% of this is tokenization with the Llama-2
tokenizer; building the suffix array takes 9\,s. The index is 13\% larger than
Lucene's.

\subsection{Query Latency}
On MS~MARCO, the first stage of InfiniSearch takes 32.7\,s per query (median;
p90 68.2\,s), against 0.39\,s for BM25 (Table~\ref{tab:efficiency}). 81\% of
this time is spent in index calls: constructing the CNF queries, which counts
candidate terms and their combinations, accounts for 48\% (median 13.1\,s), and
executing them for 33\% (10.3\,s). On the 32.9B-token index, each of the
roughly 227 calls per query takes 155\,ms on average, so latency is driven by
query construction (up to 179\,s for the slowest query) rather than by the pool,
which the budget $B$ caps at about 25k documents. Resolving the pool to text
takes the remaining 19\% (8.3\,s, 0.33\,ms per document), and the bi-encoder
another 20.8\,s. On the much smaller BEIR corpora, index calls are cheap
(15--161\,ms per query in total), and resolving dominates, at a cost linear in
the pool size (Appendix~\ref{ap:ap_efficiency}).

The latency is considerably worse than for a Lucene-based index for the tested sized. We stress that there is no Lucene-based index of pretraining size available for comparison.
The findings point to CNF-query construction as the main lever for efficiency gains.
Constructing queries is costly in itself, since the heuristic counts candidate
terms and combinations before choosing which queries to run.
More importantly, query precision determines the pool size and thus the cost of
every later stage. This suggests that future work should focus on constructing more precise CNF queries to make \projName{} both faster and more effective.

\section{Retrieval at Scale: A Qualitative Analysis}
\label{sec:qual}

Since there are currently no retrieval benchmarks on an LLM-size corpus, we conduct a case study on the OLMO2-index to illustrate what additional insights we can add to OlmoTrace \cite{liuOLMoTraceTracingLanguage2025}. OlmoTrace was developed as a companion to Olmo 2 \cite{olmo2OLMo22025}, which was trained on a 4.6T token dataset requiring over 22TB of disk space. OlmoTrace aims to connect the training data to model outputs, but it is limited to documents that contain a span matching the model output verbatim. \projName{} removes that limitation.

\begin{table*}[th!]
\centering
\footnotesize
\setlength{\tabcolsep}{4pt}
\begin{tabular}{p{0.10\textwidth} p{0.84\textwidth}}
\toprule
\textbf{Query} &
What is vicarious trauma and how to cope with it? \\
\midrule

\textbf{Response} &
It is not the same as experiencing trauma firsthand, but it can
\tracehl{lead to similar symptoms such as anxiety, depression,
burnout, and} changes in outlook on life. \\
\midrule

\multirow{4}{*}{\textbf{Traces}}
&
\dots both \colorbox{red!25}{nicotine and caffeine}
\tracehl{lead to similar symptoms such as anxiety},
headache, insomnia, nausea, vomiting, dizziness,
and arrhythmia \dots
\\

&
\dots \colorbox{red!25}{SOC} has been associated with
negative psychological
\tracehl{symptoms such as anxiety, depression, burnout,}
anger \dots
\\

&
\dots \colorbox{yellow!25}{Doctors' mental disorders},
\tracehl{such as anxiety, depression, burnout, and}
exhaustion, have been extensively investigated \dots
\\

&
\dots \colorbox{red!25}{perfectionism} has been linked
to a number of mental health problems,
\tracehl{such as anxiety, depression, burnout, and}
eating disorders. \dots
\\

\bottomrule
\end{tabular}
\caption{
Top OlmoTrace results for an excerpt from a query about vicarious trauma. \colorbox{green!25}{Green} highlights the match found by OlmoTrace. \colorbox{red!25}{Red} highlights an unrelated topic. \colorbox{yellow!25}{Yellow} highlights a plausibly related topic.
}
\label{tab:olmo_traces}
\end{table*}

\begin{table*}[h!]
\centering
\footnotesize
\setlength{\tabcolsep}{4pt}
\begin{tabular}{
    p{0.26\textwidth}
    p{0.50\textwidth}
    p{0.18\textwidth}
}
\toprule
\textbf{Query (generated claim)} &
\textbf{Excerpt from top-ranked document retrieved from OLMOs training corpus} &
\textbf{CNF match} \\
\midrule
Vicarious trauma is not the same as experiencing trauma firsthand
&
However, \colorbox{green!25}{[Vicarious] trauma} doesn’t always have to be experienced firsthand to cause damage.  \dots
&
\colorbox{green!25}{(vicarious trauma)} AND \colorbox{yellow!25}{(empathic engagement)} \\
Symptoms of vicarious trauma can include anxiety
&
Symptoms of \colorbox{green!25}{vicarious trauma} include:
[\dots] \colorbox{yellow!25}{hypervigilance} about personal safety \dots
&
\colorbox{green!25}{(vicarious trauma)} AND \colorbox{yellow!25}{(hypervigilance)} \\
Symptoms of vicarious trauma can include depression
&
Ongoing [\colorbox{yellow!25}{compassion fatigue}] can lead to PTSD, Depression,
Stress Related Chronic Illnesses \dots
&
\colorbox{yellow!25}{(compassion fatigue)} AND \colorbox{yellow!25}{(emotional exhaustion)} \\
Symptoms of vicarious trauma can include changes in outlook on life
&
\dots \colorbox{green!25}{vicarious trauma} changes a helper’s view about themselves, others and the world via alteration of \colorbox{yellow!25}{cognitive schemas} \dots
&
\colorbox{green!25}{(vicarious trauma)} AND \colorbox{yellow!25}{(cognitive schemas)} \\
Symptoms of vicarious trauma can include burnout
&
\dots advocates can experience \colorbox{red!25}{compassion satisfaction}
and vicarious resilience, which can help mitigate
the effects of \colorbox{green!25}{burnout} \dots
&
\colorbox{yellow!25}{(compassion fatigue)} AND \colorbox{yellow!25}{(emotional exhaustion)} \\
\bottomrule
\end{tabular}
\caption{
Excerpt from the top-ranked result that \projName{} retrieved from the training corpus of Olmo-2:32B the claims from the OlmoTrace passage. \colorbox{green!25}{Green} highlights indicate matches to entities from the query while \colorbox{yellow!25}{yellow} highlights indicate matches to keyword expansions. 
\colorbox{red!25}{Red} highlights a fuzzy match to the query. 
}
\label{tab:search_traces}
\end{table*}

\subsection{Comparison to OlmoTrace}

In this case study, we first prompt Olmo-2:32B with the query
about \textit{vicarious trauma}, shown in \autoref{fig:figure_1}.
We then collect the model output (full output is available in \cref{ap:olmo-out}), and the traces that OlmoTrace\footnote{Olmo Models and OlmoTrace are currently hosted by AI2 at \url{https://playground.allenai.org/}.} provides for the model output. OlmoTrace returned matches for segments of the output based on the token unigram probability of maximal overlapping spans, further processed by BM25 for relevance estimation.

Table \ref{tab:olmo_traces} illustrates the matches to  OLMOs training corpus that OlmoTrace provided for a part of the response. We can see, that although the found documents contain the verbatim match to the list of symptoms, it comes from the contexts about vastly different causes, not clearly relevant to the actual query. This is problematic if e.g. the sources are needed for fact-checking, to corroborate the claims in the model output. 

In \projName{}, we first decompose a paragraph from the model output into eleven claims,\footnote{This step is not part of \projName{} pipeline, as it is necessarily application-specific, and depends on the additional constraints about what kinds of spans need to be extracted, and at what computational budget. In this illustration, we assumed a scenario of claim fact-checking, and used Claude-Opus-4.6 with manual verification for claim extraction. In real applications, more lightweight options targeted to specific kinds of information could be selected based on the specific use case.} which are used individually as queries, with \infinigram{} API as the underlying index. We limited the number of documents to 300 to avoid latency issues. After deduplication, we retrieved 179 documents per query on average. For each query, we rerank the retrieved documents as described in \cref{sec:retr}. 

To compare the \projName{} results with the original OlmoTrace documents for relevance to the query, we include the traces from OlmoTrace in the reranking step to see where they would be ranked in the list of most relevant documents. If the OlmoTrace documents already ranked high, \projName{} would confer little additional benefit. However, we find that the average best rank for a document from OlmoTrace was 84.5 (or 46.9 when normalized to 100), and  %
the best overall rank of any document from OlmoTrace was 46. This suggests that on average, the documents retrieved with \projName{} are more relevant to the model output.

We further analyze the top results qualitatively, and we confirm that the generated claims can generally be corroborated with the top-ranked document. \autoref{tab:search_traces} shows the top ranked document for each claim extracted from the corresponding passage in \autoref{tab:olmo_traces}.
For all claims, the top document was found using keyword expansion, which shows that it does help to find highly relevant documents. 

\subsection{Error Analysis}
While we find the top-ranked document mostly highly relevant as a source of corroborative evidence for the model claims, we also considered the top-20\footnote{Top-20 rankings are provided in the appendix (\ref{ap:qualan}). Top ranking for other claims are provided in \autoref{tab:search_traces_long}, and full rankings are available in the project repository.} documents for the claims in this case study. We find that for most claims, irrelevant documents were also included in the top 20. We identify two main sources for this:
\begin{itemize*}
    \item \textbf{Standalone queries}. For example, in the claim regarding \textit{anxiety}, a standalone query for the associated term \textit{cognitive disruption} produced a low number of matches in the index and was included for retrieval. In this way we retrieved a document about concussions, not relevant for the main query but ranked 20. The same can happen with key-term expansions: e.g. \textit{empathic strain}, a key-term expansion for \textit{vicarious trauma}, also retrieves documents about irrelevant types of strains.
    \item \textbf{Irrelevant AND-queries}. The combination of AND-queries that do not include the first, most specific key term, often leads to irrelevant results. These were often below the top 20 results. Examples include {\tt (symptomatology) AND (primary trauma)} (top 16) and {\tt (warning signs) AND (anxiety)} (top 30).   
\end{itemize*}

These errors show that while the standard pipeline that we have used throughout our experiments works well without hyperparameter tuning, there is still room for improvement, especially when retrieving more documents. We estimate the quality of a query based on the estimated matches in the corpus as a proxy for specificity. Future work could include a priori estimation of the relevance of each query, to improve the precision of the first-stage retrieval. This could be done with term weighting or other relevance estimations.

\section{Conclusion and Future Work}
We presented \projName{}, a novel first stage retriever based upon byte-suffix-arrays (BSA), uniquely positioned to enable retrieval at LLM training data scale. We showed that it is possible to overcome the BSA limitation of verbatim matches via LLM-based keyword expansion. %
We established that our system performs well as a retriever in common retrieval benchmarks, and we also illustrated how it can be used on the scale of a pretraining corpus to conduct dataset analysis that is not at all possible with other retrievers. We hope that \projName{} could unlock new directions for research on data attribution, generalization, and benchmark contamination, based on the fully open-source LLMs such as OLMo. 

The current implementation of \projName{} is intended as a proof-of-concept, and we expect that the pipeline could be significantly improved for specific research goals. The directions for that include prompt engineering for keyword expansion (targeting specific information needs), fine-tuning, or using more modern embedders, such as FollowIR \cite{wellerFollowIREvaluatingTeaching2025} as well as large cross-encoders, such as RankT5 \cite{zhuangRankT5FineTuningT52022}, or LLM-based list-wise rerankers \cite{maZeroShotListwiseDocument2023}. 

Future work could further improve retrieval performance. At the index level, a newer version of \infinigram{} developed by \citet{xuInfinigramMiniExact2025} could further reduce hardware requirements. It is based on an FM-Index \cite{ferraginaOpportunisticDataStructures2000}. FM-Indices offer smaller index sizes through compression while maintaining the scalability and retrieval capabilities of BSAs. Our contribution (mitigating the limitation to verbatim matches) is compatible with using FM-indices. Future work should also investigate more sophisticated methods for CNF-query construction, as the latency bottleneck comes mainly from the number of queries that have to be executed to retrieve results. Another promising direction is suggested by {\em Deep Retriever} \cite{jiangDeepRetrievalHackingReal2025}, a pipeline for training a model for keyword expansion using reinforcement learning. This could mitigate the main driver of latency, by constructing queries in one pass without using heuristics to check many potential queries for their precision. It would also make the system independent of an external LLM for keyword expansion.

\section*{Limitations}

\paragraph{Dependence on external resources.} \projName{} is based on \infinigram{}, which is currently the only publicly available resource of its kind. It was made for a specific dataset of a specific English LLM, OLMo 2 \cite{olmo2OLMo22025}. 
While it is theoretically possible to build such indices for any corpus, for labs with constrained resources, the utility of \projName{} necessarily relies on existence of such indices for the model of interest. However, this applies to any work on LLM-scale retrievers. There are few high-performing models with open training data, and even fewer indices for them. Work on both is crucial for key unresolved challenges in LLM research, including the assessment of generalization, test contamination, data representativeness, and data attribution. 

\paragraph{Efficiency.} The main goal of \projName{} pipeline is to support retrieval at LLM-scale corpora. When the same approach is used at the scale of much smaller retrieval benchmarks, \projName{} is significantly slower than a standard Lucene-based first-stage retrieval (see \autoref{sec:efficiency}). To some extent this can still be mitigated by parallelization and optimized sharding, similar to the official \infinigram{} setting. Additionally, better query constructions could improve the number of documents that need to be retrieved for satisfactory retrieval. But this is an inconvenience for the development stage (running retrieval evaluations), rather than the intended use of the system at LLM-scale.
Ideally, the comparison between Lucene-Based indices and \projName{} should be scaled to pretraining size. This is however prohibitively costly.

\paragraph{Brittleness.} The retriever works with exact matching of tokens and differentiates between tokens at the start of the word, e.g., {\tt Great} versus {\tt "Great}. The retrieval is therefore brittle to misspellings and to artifacts preceding terms. This is a general problem for LLM corpus preprocessing, which retrieval systems inherit.

\paragraph{Test contamination.} Finally, one limitation of all pipelines with LLM-based keyword expansion is the unknown overlap between evaluation datasets and training datasets. LLMs might have an unfair advantage if they have implicit knowledge about the most relevant documents from the datasets. However, to even establish whether that is the case, and more broadly consider the issues of data contamination, we need systems like \projName{}.

\section*{Ethical considerations}

This work was performed as an academic research project, under applicable data processing exemptions for research purposes. We relied on the following resources and benchmarks: Beir \cite{thakurBEIRHeterogenousBenchmark2021}, MS Marco \cite{pradeepRagnarokReusableRAG2024}, InfiniGram \cite{liuInfinigramScalingUnbounded2024}, to which we refer for the license information. No human subject data has been collected or released in scope of this work. 

InfiniGram \cite{liuInfinigramScalingUnbounded2024} is based on the data of Olmo2 \cite{olmo2OLMo22025}, to which we refer for the legal grounds for data collection, and discussion of steps taken to avoid inclusion of individually identifying information. The analysis conducted in scope of this work focused on general information retrieval scenarios, and did not target such information.

\projName{} establishes the theoretical possibility of retrieval over LLM-scale corpora beyond verbatim matching. Such a technology could assist not only researchers studying generalization, but also support broader data attribution efforts, which could be beneficial for both creators whose data was included in the training corpora, and LLM users \cite{WuhrlRuckdeschelEtAl_2026_Human-Centric_Framework_for_Data_Attribution_in_Large_Language_Models}. However, this hinges on the currently-absent transparency about training data on the part of commercial LLM services. In the absence of clear legal protections for non-commercial projects, and their clear separation from commercial services, the efforts towards data transparency (and especially facilitating access via public search interfaces) could have the unintended effect of opening them to legal action, overall aimed at the commercial services.

\section*{Acknowledgements}

This work was supported by a research grant (VIL60860) from Villum Fonden. We are grateful for valuable feedback from our colleagues, including Christian
Hardmeier, Leonie Weissweiler, and Alexander Koller.
Generative AI was not used in writing of this work. 
Claude-Opus-4.6 was used for coding assistance, with manual verification.

\bibliography{better_indexing,custom}

\appendix

\section{Appendix}
\label{sec:appendix}

\subsection{Prompt Engineering and Keyword expansion}
\label{ap:prompt}
There were many iterations on the final prompt used to generate the keywords. 
We settled on a prompt that limits the number of keywords that are generated. If not specified, the LLM will often generate many unspecific keywords. We included an example for the kind of expansion that we are interested in.  An earlier version of the prompt has also returned a list of verbs. These were not helpful for query generation and were ultimately not used in the pipeline.
The Final prompt can be seen in 
\autoref{fig:query_prompt}:
\begin{figure*}[t]
\centering
\begin{promptbox}
\textbf{SYSTEM:}
You are a  keyword expansion expert. The keywords need to be designed, so that they work with a search system that can only find verbatim matches of keywords, and their logical expansion. You can for example search for ('Waterpark' AND 'Attractions') to find all documents, where both terms occur. The challenge is to find keywords that work for such a search system. They need to be specific enough to find only relevant documents, but expansive enough to find clearly relevant documents with terminology that differs from the query.

Given a query,  find relevant search terms in three categories:

1.KEY\_ENTITIES: Identify the key entities in the query. They describe the information need of query. These key entities should be non-overlapping noun phrases. For each of these terms, do a keyword expansion.
Find the top five most important terms that could replace the key entity in a document. Keep the specific lexical search engine in mind. These terms should help find relevant documents with little lexical overlap with the query.

2. ASSOCIATED\_TERMS: Terms that don't fit to any entity but are relevant to the entire query. Think about good individual key phrases for finding relevant documents to the query that might have no overlap with the key entities. Find the top ten most impactful additional terms for the query. Specific terms and references are better than generic terms. 

Format: JSON object. KEY\_ENTITIES keys contain term lists. An "ASSOCIATED" key contains topic-level terms. Order by specificity. Every term must be a phrase (1-4 words) that literally appears in real documents. Return ONLY valid JSON.
\end{promptbox}
\caption{Prompt used for query decomposition.}
\label{fig:query_prompt}

\end{figure*}
We tested several other models  for the keyword expansion, but found them not complying with the instructions to a sufficient degree.
In our tests, the smaller open models, such as qwen3.6:0.8B or Gemma 4:4B did not perform the task successfully. While most models, with the exception of smollm2:135M \cite{allalSmolLM2WhenSmol2025}, are able to structure the output accordingly, the keywords they generate do not benefit the search task. 
Other models, including qwen 3.6 \cite{qwen} or deepseek-v4 \cite{deepseek}, only returned valid JSON for a fraction of requests.

\subsection{Query construction}
\label{ap:ap_queries}

The amount of queries rises with the number of keywords that are extracted. with $k$ keywords, constructing all $r$-way combinations is given by the following formula:
\begin{equation}
6^r \binom{k}{r} + 10 \cdot 6^{r-1} \binom{k}{r-1}
\label{eq:combinations}
\end{equation}
Commonly on the MSMarco-dataset, there are either two or three keywords extracted. 21 out of 301 topics produce four keywords, and only one topic produces five.\\
For $k=2$, there are 360 three-way combinations of keywords and 156 two-way-combinations. \\
For $k=3$, there are 1296 three-way combinations of keywords and 288 two-way-combinations.

\subsection{Details on retrieval performance}
We report additional retrieval metrics for our runs which were not reported for the compared baselines in \autoref{tab:retrieval_results_full}. These metrics are commonly reported but were not reported by in the \cite{kamalloo:2024}.
\label{sec:retrieval_details}
\begin{table*}[t]
\centering
\begin{tabular}{@{}cccccc@{}}
\toprule
               & \multicolumn{5}{c}{\textbf{Benchmark}}                               \\ 
\textbf{Metric}         & TREC RAG 2024 & FEVER & Climate-Fever & TREC-Covid & SciFact \\ \cmidrule(l){1-1} \cmidrule(l){2-6} 
NDCG@10        & 0.54          & 0.84  & 0.25          & 0.746      & 0.7     \\
MRR            & 0.90          & 0.84  & 0.35          & 0.89       & 0.67    \\
MAP            & 0.16          & 0.8   & 0,19          & 0.09       & 0.62    \\
R@100          & 0.22          & 0.96  & 0.56          & 0.13       & 0.89    \\
R@Pool         & 0.489         & 0.97  & 0.69          & 0.58       & 0.92    \\ \bottomrule
\end{tabular}
\caption{Retrieval results on four benchmarks of the BEIR benchmark suite and the TREC RAG 2024 challenge. R@Pool gives the total recall that the \textbackslash{}projname achieves on the pool of retrieved documents. All results were achieved without hyperparameter-tuning or preprocessing. }
\label{tab:retrieval_results_full}
\end{table*}

\subsection{Efficiency Information}
\label{ap:ap_efficiency}

\begin{table}[t]
\centering
\small
\setlength{\tabcolsep}{4pt}
\begin{tabular}{lrrrrr}
\toprule
 & & \multicolumn{2}{c}{Build time (s)} & \multicolumn{2}{c}{Size (MB)} \\
\cmidrule(lr){3-4}\cmidrule(lr){5-6}
Corpus & Docs & IG & BM25 & IG & BM25 \\
\midrule
NFCorpus & 3.6k &  3.2 &  1.4 &  8.1 &   7.7 \\
SciFact  & 5.2k &  3.4 &  1.5 & 10.7 &  10.6 \\
ArguAna  & 8.7k &  3.5 &  1.8 & 11.5 &  13.0 \\
FiQA     &  58k &  5.8 &  3.8 &   72 &    67 \\
COVID    & 171k & 15.1 & 11.0 &  316 &   263 \\
Touch\'e & 383k & 48.7 & 35.1 & 1010 &   891 \\
\bottomrule
\end{tabular}
\caption{Index construction (median of three builds). IG: infini-gram; BM25:
Lucene index with positions and stored documents. The $\sim$3\,s minimum for IG
is start-up overhead. COVID: TREC-COVID.}
\label{tab:index-build}
\end{table}

\begin{table}[t]
\centering
\small
\setlength{\tabcolsep}{2.2pt}
\begin{tabular}{cccccccc}
\toprule
 & BM25 & \multicolumn{3}{c}{InfiniSearch} & & \multicolumn{2}{c}{Recall@pool} \\
\cmidrule(lr){2-2}\cmidrule(lr){3-5}\cmidrule(lr){7-8}
Corpus & med. & med. & p90 & index & Pool & IS & BM25 \\
\midrule
NFCorpus &  5 &   36 &  149 &  15 &   338 & .383 & .329 \\
SciFact  & 21 &   56 &  142 &  26 &   337 & .916 & .931 \\
ArguAna  & 25 &   81 &  154 &  41 &   527 & .920 & .967 \\
FiQA     & 25 &  298 &  901 &  34 &  3.9k & .781 & .809 \\
COVID    & 27 & 1980 & 2490 & 161 & 16.5k & .712 & .779 \\
Touch\'e & 32 & 2100 & 4220 &  76 &  8.4k & .881 & .956 \\
\bottomrule
\end{tabular}
\caption{First-stage latency per query (ms; laptop, warm cache). BM25: median
for top-1000 including reading the documents. InfiniSearch: median, 90th
percentile, and mean time for constructing and executing CNF queries (index);
the rest is resolving documents. Pool: mean documents retrieved. Recall@pool:
InfiniSearch pool vs.\ BM25 at the same depth.}
\label{:ap_efficiency}
\end{table}

\subsection{Hyperparameters for the pipeline}
\autoref{tab:hyperparams} shows the hyperparameters that are necessary to configure the pipeline and shows the default values that we have chosen.

\begin{table*}[t]
\centering
\small
\begin{tabular}{
@{}
>{\raggedright\arraybackslash}p{0.22\textwidth}
>{\raggedright\arraybackslash}p{0.52\textwidth}
>{\raggedright\arraybackslash}p{0.18\textwidth}
@{}
}
\toprule
Name & Description & Default \\ \midrule

max\_standalone\_key &
Maximum count under which standalone key terms and expansions are used for retrieval &
5000 \\

max\_standalone\_assoc &
Maximum count under which associated terms are used for retrieval &
1000 \\

prox\_peek &
Maximum distance between AND-parts for a CNF-query when counting standalone CNF-pieces. This is passed to \textit{max\_diff\_tokens} in the infinigram call. &
10 \\

prox\_cross &
Maximum distance between AND-parts for a CNF-query when counting combined queries. This is passed to \textit{max\_diff\_tokens} in the infinigram call. &
200 \\

max\_clause\_freq &
Maximum amount of matches above which infinigram samples instead of returning accurate retrieval results passed through to infinigram on every call. &
8000000 \\

max\_budget &
Maximum amount of counts that all CNF-queries for retrieval can have. &
\\

max\_doc\_len &
Number of tokens retrieved for each document. If the document is short, retrieve the entire document. This is counted from the beginning of the document. &
20000 \\

biencoder\_model &
Model used as biencoder &
sentence-transformers all-MiniLM-L6-v2 \\

biencoder\_top\_k &
How many documents are returned from semantic filtering &
1000 \\

crossencoder\_model &
Model used as crossencoder &
cross-encoder ms-marco-MiniLM-L-12-v2 \\

crossencoder\_top\_k &
How many documents are returned from reranking &
100 \\

\bottomrule
\end{tabular}

\caption{
Hyperparameters and defaults for the retrieval pipeline.
Default hyperparameters for the biencoder and crossencoder are not included.
}
\label{tab:hyperparams}
\end{table*}

\subsection{Qualitative Analysis}
\subsubsection{Output from Olmo}
\autoref{fig:olmoout} shows the full output from Olmo2:32B for the query from \autoref{fig:figure_1}.
\label{ap:olmo-out}

\begin{figure*}[t]
\centering
\begin{promptbox}
    Vicarious trauma refers to the psychological impact on an individual who is exposed to the traumatic experiences of others through empathic engagement, such as therapists, emergency service workers, health care professionals, or even those watching graphic news reports. It is not the same as experiencing trauma firsthand, but it can lead to similar symptoms such as anxiety, depression, burnout, and changes in outlook on life.

Here are some ways to cope with vicarious trauma:

    Self-awareness: Being aware of the signs and symptoms of vicarious trauma in yourself, such as changes in mood, sleep patterns, and increased irritability, is the first step towards coping with it.

    Professional supervision and debriefing: Participating in structured supervision or debriefing sessions with colleagues or supervisors can provide support and perspective on the challenges you're facing.

    Limit exposure: Establish healthy boundaries and limit your exposure to traumatic material when possible. Breaks from the content can help prevent emotional overload.

    Self-care: Prioritize activities that help maintain your physical and mental health, such as exercise, healthy eating, and getting enough sleep.

    Seek support: Lean on friends, family, and trusted colleagues for emotional support. Joining a support group with others who understand vicarious trauma can also be beneficial.

    Professional therapy: If symptoms persist, seeking professional help from a psychologist or counselor trained in trauma can provide you with the necessary tools to heal and manage the impact of vicarious trauma.

    Resilience building: Engage in activities that build resilience and a sense of purpose, such as mindfulness, meditation, or hobbies that bring you joy.

    Boundaries in the workplace: Establish clear boundaries with clients or in your professional role, ensuring that you have time away from vicarious trauma exposure.

    Limit media consumption: Be mindful of your media consumption. Watching too much graphic or distressing news can contribute to vicarious trauma.

    Continued education: Keep learning about vicarious trauma and effective coping strategies. This can empower you with knowledge and tools to manage your experiences.

Remember that coping with vicarious trauma is an ongoing process, and it's important to be patient with yourself as you navigate through these experiences. It's also crucial to recognize that seeking help is a sign of strength, not weakness.
\end{promptbox}
\caption{Olmo2 Output for the query from \autoref{fig:figure_1}.}
\label{fig:olmoout}
\end{figure*}

\label{ap:qualan}

\subsection{Top results for all extracted claims of the passage.}

We show the top ranked results of all eleven claims that were extracted from \autoref{fig:claim-passasge} in \autoref{tab:search_traces_long}. Here we see that in addition to full matches, we also can have fuzzy matches of parts of the CNF-queries. This is due to the decomposition of two-term keywords when creating cnf-queries.
\begin{table*}[h!]
\centering
\footnotesize
\setlength{\tabcolsep}{4pt}
\begin{tabular}{
    p{0.26\textwidth}
    p{0.50\textwidth}
    p{0.18\textwidth}
}
\toprule
\textbf{Query (generated claim)} &
\textbf{Excerpt from top-ranked document retrieved from OLMOs training corpus} &
\textbf{CNF match} \\
\midrule
Emergency service workers can experience vicarious trauma
&
 \colorbox{green!25}{Vicarious trauma} can happen to anyone, but  \colorbox{yellow!25}{first responders} and counselors themselves have an increased risk 
&
 \colorbox{yellow!25}{(first responders)} AND  \colorbox{green!25}{(vicarious trauma)} \\
watching graphic news reports can cause vicarious trauma
&
\ldots[Vicarious Trauma] can also affect civilians who [\ldots] are deeply impacted by stories that they are exposed to (maybe \colorbox{red!25}{graphic news accounts}, \ldots \ldots
&
\colorbox{green!25}{graphic news reports} \\
Healthcare professionals can experience vicarious trauma
&
\colorbox{green!25}{Vicarious trauma} among \colorbox{red!25}{healthcare professionals}, and other individuals in the helping profession, has been magnified over the past year. \ldots
&
\colorbox{yellow!25}{(healthcare workers)} AND \colorbox{green!25}{(vicarious trauma)} \\
First responders can experience vicarious trauma
&
\colorbox{green!25}{Vicarious trauma} can happen to anyone, but \colorbox{green!25}{first responders}
and counselors themselves have an increased risk \dots
&
\colorbox{green!25}{(first responders)} AND (\colorbox{green!25}{vicarious trauma)} \\
Symptoms of vicarious trauma can include anxiety
&
Symptoms of \colorbox{green!25}{vicarious trauma} include:
[\dots] \colorbox{yellow!25}{hypervigilance} about personal safety \dots
&
\colorbox{green!25}{(vicarious trauma)} AND \colorbox{yellow!25}{(hypervigilance)} \\
Symptoms of vicarious trauma can include depression
&
Ongoing [\colorbox{yellow!25}{compassion fatigue}] can lead to PTSD, Depression,
Stress Related Chronic Illnesses \dots
&
\colorbox{yellow!25}{(compassion fatigue)} AND \colorbox{yellow!25}{(emotional exhaustion)} \\
Symptoms of vicarious trauma can include changes in outlook on life
&
\dots \colorbox{green!25}{vicarious trauma} changes a helper’s view about themselves, others and the world via alteration of \colorbox{yellow!25}{cognitive schemas} \dots
&
\colorbox{green!25}{(vicarious trauma)} AND \colorbox{yellow!25}{(cognitive schemas)} \\
Symptoms of vicarious trauma can include burnout
&
\dots advocates can experience \colorbox{red!25}{compassion satisfaction}
and vicarious resilience, which can help mitigate
the effects of \colorbox{green!25}{burnout} \dots
&
\colorbox{yellow!25}{(compassion fatigue)} AND \colorbox{yellow!25}{(emotional exhaustion)} \\
Therapists can experience vicarious trauma
&
Therapists and \colorbox{yellow!25}{counselors} \dots may be vulnerable
to \colorbox{green!25}{vicarious trauma}.
&
\colorbox{yellow!25}{(counselors)} AND \colorbox{green!25}{(vicarious trauma)} \\
Vicarious trauma is not the same as experiencing trauma firsthand
&
However, trauma doesn’t always have to be experienced
firsthand to cause damage \dots
&
\colorbox{green!25}{(vicarious trauma)} AND \colorbox{yellow!25}{(empathic engagement)} \\
Vicarious trauma refers to the psychological impact on an individual who is exposed to the traumatic experiences of others through empathic engagement
&
\dots constant and intense[\ldots] \colorbox{red!25}{exposure to trauma} with little or no variation in work task \dots
&
\colorbox{green!25}{(vicarious trauma)} AND \colorbox{yellow!25}{(trauma exposure)} \\
\bottomrule
\end{tabular}
\caption{
Excerpt from the top-ranked result that \projName{} retrieved from the training corpus of Olmo-2:32B for all extracted queries. \colorbox{green!25}{Green} highlights indicate matches to entities from the query while \colorbox{yellow!25}{yellow} highlights indicate matches to keyword expansions. No associated term match was included in the top result for any query. \colorbox{red!25}{Red} highlights a fuzzy match to the query. 
}
\label{tab:search_traces_long}
\end{table*}
'

\subsection{Top 20 Results for \projName{} claim queries}
We conducted the claim extraction for the paragraph in \autoref{fig:claim-passasge} that exceeded the small section for which OlmoTrace was able to find a match in the training data.
We have created eleven queries for \projName{} for each claim in the passage from which the OlmoTrace excerpt was selected. The top 20 results from \projName{} can be seen in tables \ref{tab:first_rank_table} through \ref{tab:last_rank_table}.
\begin{figure*}[t]
\centering
\begin{promptbox}
Vicarious trauma refers to the psychological impact on an individual who is exposed to the traumatic experiences of others through empathic engagement, such as therapists, emergency service workers, health care professionals, or even those watching graphic news reports. It is not the same as experiencing trauma firsthand, but it can lead to similar symptoms such as anxiety, depression, burnout, and changes in outlook on life.
\end{promptbox}
\caption{passage from which claims were extracted}.
\label{fig:claim-passasge}
\end{figure*}
The tables have been deduplicated before the final ranking. The full tables can be found in the supplementary materials attached to the submission.
\onecolumn
\small
\begin{longtable}{>{\raggedright\arraybackslash}p{0.06\linewidth}>{\raggedright\arraybackslash}p{0.68\linewidth}>{\raggedright\arraybackslash}p{0.20\linewidth}}
\caption{Top 20 results for the claim Emergency service workers can experience vicarious trauma} 
\label{tab:first_rank_table}
\\
\toprule
Rank & text & from\_queries \\
\midrule
\endfirsthead
\toprule
Rank & text & from\_queries \\
\midrule
\endhead
\midrule
\endfoot
\bottomrule
\endlastfoot
1 & Seeing and experiencing disasters and traumatic events with one’s own eyes can be a life-altering experience. Images of disaster and loss can stick with a person, creating intrusive thoughts and causing a person to seek refuge from their mental state. The effects of this trauma can infiltrate every aspect of a person’s life, compromising one’s mental and emotional health. It can also carry several social or professional hardships and may even develop into post-traumatic stress disorder (PTSD). \ldots & (first responders) AND (vicarious trauma) \\
2 & On any given day, a firefighter, paramedic or police officer can see more violence and trauma than the average civilian may encounter in their entire lifetime.  The result of which, research suggests, may lead to damaging psychological effects. This is something we don't always consider when we see our friends and family who serve in these jobs.  Moreover, there is a large contingent of volunteers who serve our communities in these roles. Many people sign up to make a difference and serve their \ldots & (first responders) AND (compassion fatigue); (first responders) AND (vicarious trauma) \\
3 & The role of Filipino masculine ideology on the adaptive coping, psychological wellbeing and vicarious trauma of first responders  Document Type   Publication Date    The exposure of first responders to emergencies and disaster puts them at high risk of experiencing vicarious trauma. First responders are typically male and literature suggests that a negative connotation of masculinity creates a stigma that leads to low help-seeking behaviour and poorer psychological outcomes. This study examines \ldots & (first responders) AND (vicarious trauma); (first responders) AND (vicarious trauma) \\
4 & Friday , June 18 2021  Vicarious Trauma For Doctors And Nurses Who Treat Victims Of Gun Violence: Shots  Clinicians and first responders who care for the gun violence can suffer what's called vicarious trauma.  Image Source / Getty Images  hide caption  toggle caption  Image Source / Getty Images   Image Source / Getty Images  Gun violence has become a part of everyday life in America and the work lives of doctors, nurses and first responders, too.  After the National Rifle Association told \ldots & (first responders) AND (secondary trauma) \\
5 & © 2021 KMUW Play Live Radio Next Up: Available On Air Stations  How Doctors And Nurses Cope With The Human Toll Of Gun Violence  Clinicians and first responders who care for victims of gun violence can suffer what's called vicarious trauma.  Gun violence has become a part of everyday life in America and of the work lives of doctors, nurses and first responders, too.  After the National Rifle Association told doctors to "stay in their lane" in response to a policy proposal from the American \ldots & (first responders) AND (secondary trauma) \\
6 & How Doctors And Nurses Cope With The Human Toll Of Gun Violence  Clinicians and first responders who care for victims of gun violence can suffer what's called vicarious trauma. /Image Source/Getty Images  Gun violence has become a part of everyday life in America and of the work lives of doctors, nurses and first responders, too.  After the National Rifle Association told doctors to "stay in their lane" in response to a policy proposal from the American College of Physicians for reducing \ldots & (first responders) AND (secondary trauma) \\
7 & Play Live Radio Next Up: Available On Air Stations  How Doctors And Nurses Cope With The Human Toll Of Gun Violence  Clinicians and first responders who care for victims of gun violence can suffer what's called vicarious trauma. Image Source/Getty Images  Gun violence has become a part of everyday life in America and of the work lives of doctors, nurses and first responders, too.  After the National Rifle Association told doctors to "stay in their lane" in response to a policy proposal from the \ldots & (first responders) AND (secondary trauma) \\
8 & How Doctors And Nurses Cope With The Human Toll Of Gun Violence  Clinicians and first responders who care for victims of gun violence can suffer what's called vicarious trauma. Credit / Image Source/Getty Images  Gun violence has become a part of everyday life in America and of the work lives of doctors, nurses and first responders, too.  After the National Rifle Association told doctors to “stay in their lane” in response to a policy proposal from the American College of Physicians for \ldots & (first responders) AND (secondary trauma) \\
9 & How Doctors And Nurses Cope With The Human Toll Of Gun Violence  Clinicians and first responders who care for victims of gun violence can suffer what's called vicarious trauma. (Image Source/Getty Images)  Gun violence has become a part of everyday life in America and of the work lives of doctors, nurses and first responders, too.  After the National Rifle Association told doctors to "stay in their lane" in response to a policy proposal from the American College of Physicians for reducing \ldots & (first responders) AND (secondary trauma) \\
10 & News 88.9 KNPR Classical 89.7 KCNV  member station  Shots - Health News  How Doctors And Nurses Cope With The Human Toll Of Gun Violence   Clinicians and first responders who care for victims of gun violence can suffer what's called vicarious trauma. Image Source/Getty Images   Gun violence has become a part of everyday life in America and of the work lives of doctors, nurses and first responders, too.  After the National Rifle Association told doctors to "stay in their lane" in response to a \ldots & (first responders) AND (secondary trauma) \\
11 & Radio Website Header-Waves 6 3.0.jpg Play Live Radio Next Up: Available On Air Stations  How Doctors And Nurses Cope With The Human Toll Of Gun Violence  Clinicians and first responders who care for victims of gun violence can suffer what's called vicarious trauma.    How do doctors and nurses cope with their regular encounters with the human toll of gun violence? How does exposure to trauma affect them?  Trauma surgeon Stephanie Bonne estimates that she and her colleagues at University \ldots & (first responders) AND (secondary trauma) \\
12 & First Responders\& PTSD  Photo by Ibrahim Mushan,  First responders and military veterans are a unique class of people in that they frequently face life or death circumstances that can cause vicarious trauma, and over time, full PTSD. In fact, a recent study by University of Phoenix found that one-third of first responders studied had been formally diagnosed with depression or PTSD. 85\% of them had experienced mental health symptoms.  In recent years, a greater push toward trauma education and \ldots & (first responders) AND (vicarious trauma) \\
13 & High Risk Humans   Vicarious Trauma: The Effects onFirst Responders  By Damedrist Wright, MS SFC (RET)  As our lives and lifestyles become increasingly complex and hectic, resilience has become an ever more important skill to have. First responders who are exposed to disruptive events involving death, violence, or life-threatening situations on a daily basis are in special need of this valuable trait. When traumatic events occur, the impact can have an overwhelming effect on a first responder \ldots & (first responders) AND (compassion fatigue); (first responders) AND (secondary traumatic stress) \\
14 & High Risk Humans   Vicarious Trauma: The Effects onFirst Responders  By Damedrist Wright, MS SFC (RET)  As our lives and lifestyles become increasingly complex and hectic, resilience has become an ever more important skill to have. First responders who are exposed to disruptive events involving death, violence, or life-threatening situations on a daily basis are in special need of this valuable trait. When traumatic events occur, the impact can have an overwhelming effect on a first responder \ldots & (first responders) AND (secondary traumatic stress) \\
15 & © 2022 | WUWF Public Media 11000 University Parkway Pensacola, FL 32514 850 474-2787 NPR for Florida's Great Northwest Play Live Radio Next Up: Available On Air Stations  How Doctors And Nurses Cope With The Human Toll Of Gun Violence  Clinicians and first responders who care for victims of gun violence can suffer what's called vicarious trauma. Image Source/Getty Images  Gun violence has become a part of everyday life in America and of the work lives of doctors, nurses and first responders, \ldots & (first responders) AND (secondary trauma) \\
16 & Critical Incident Stress  A critical incident can be defined as any event that has a stressful impact sufficient enough to overwhelm the usually effective coping skills of an individual. Critical incidents are abrupt, powerful events that fall outside the range of ordinary human experiences. These events can have a strong emotional impact, even on the most experienced officer or deputy. Research has shown that critical incident stress affects up to 87\% of all emergency service workers at least \ldots & (emergency service workers) AND (critical incident stress) \\
17 & Critical Incident Stress   A critical incident can be defined as any event that has a stressful impact sufficient enough to overwhelm the usually effective coping skills of an individual. Critical incidents are abrupt, powerful events that fall outside the range of ordinary human experiences. These events can have a strong emotional impact, even on the most experienced officer or deputy. Research has shown that critical incident stress affects up to 87\% of all emergency service workers at least \ldots & (emergency service workers) AND (critical incident stress) \\
18 & Government of New Brunswick  A program for New Brunswick's Emergency First Responders  What is a critical incident?  A critical incident is a situation that is traumatic for the emergency first responders involved. The incident causes the worker to experience an unusually strong emotional reaction either during or after the incident.  Examples of critical incidents include:    • serious injury or death of a co-worker in the line of duty;   • loss of a patient after prolonged rescue attempts; \ldots & (emergency medical technicians) AND (critical incident stress) \\
19 & Bringing The World Home To You  © 2021 WUNC North Carolina Public Radio 120 Friday Center Dr Chapel Hill, NC 27517 919.445.9150 | 800.962.9862 Play Live Radio Next Up: Available On Air Stations  How Doctors And Nurses Cope With The Human Toll Of Gun Violence  Clinicians and first responders who care for victims of gun violence can suffer what's called vicarious trauma.  Gun violence has become a part of everyday life in America and of the work lives of doctors, nurses and first responders, too. \ldots & (first responders) AND (secondary trauma) \\
20 & The issue of emergency service workers esw facing the critical incident stress cis  65 identifying the critical incident and wish list items therefore, the research question is what helps, hinders and might help police i conducted a study of older emergency service workers where the police to remain constant: enforce the law while facing potential tucson, az: e-s press. Critical incident stress and challenges for the emergency workplace (1) department of emergency medicine, mount sinai school \ldots & (emergency service workers) AND (critical incident stress); (emergency service workers) AND (critical incident stress); (emergency service workers) AND (critical incident stress) \\
\end{longtable}

\small
\begin{longtable}{>{\raggedright\arraybackslash}p{0.06\linewidth}>{\raggedright\arraybackslash}p{0.68\linewidth}>{\raggedright\arraybackslash}p{0.20\linewidth}}
\caption{Top 20 results for the claim Health care professionals can experience vicarious trauma} \\
\toprule
Rank & text & from\_queries \\
\midrule
\endfirsthead
\toprule
Rank & text & from\_queries \\
\midrule
\endhead
\midrule
\endfoot
\bottomrule
\endlastfoot
1 & How to Cope withVicarious Traumaas a Healthcare Professional  Watch this blog in ASL.  Vicarious trauma among healthcare professionals, and other individuals in the helping profession, has been magnified over the past year.  Vicarious Trauma Among BehavioralHealthcare Workers  Throughout their day, behavioral healthcare workers hear the stories of the people who have often experienced mild or significant trauma.  “Hearing those stories of mistreatment or difficult times can start to take an \ldots & (healthcare workers) AND (vicarious trauma) \\
2 & Compassion Fatigue  Compassion Fatigue  Medical professionals such as physicians, nurses, psychotherapists, and emergency workers, who help traumatized patients may develop their own Post Traumatic Stress Disorder (PTSD) symptoms as an indirect response to their patient’s suffering. This phenomenon has been referred to as compassion fatigue, vicarious traumatization or secondary traumatic stress.  A survey showed that “86.9\% of emergency response personnel reported symptoms after exposure to \ldots & (health care professionals) AND (secondary traumatic stress) \\
3 & Category Archives: Trauma Reactions  Compassion Fatigue  Compassion Fatigue  Medical professionals such as physicians, nurses, psychotherapists, and emergency workers, who help traumatized patients may develop their own Post Traumatic Stress Disorder (PTSD) symptoms as an indirect response to their patient’s suffering. This phenomenon has been referred to as compassion fatigue, vicarious traumatization or secondary traumatic stress.  A survey showed that “86.9\% of emergency response personnel \ldots & (health care professionals) AND (secondary traumatic stress) \\
4 & Treating Trauma Self-care for providers Bookmark and Share  Indirect Trauma  Professionals who work with trauma survivors in an open, engaged, and empathic way and who feel responsible or committed to helping them are likely to experience indirect trauma. That means that they will be transformed by the work. The way helpers understand and experience the world and themselves is changed as they enter into the world of the survivor. While trauma work can be very meaningful and rewarding, it can \ldots & empathic strain \\
5 & 013).  Prevalence Research indicates that the nature of a clinical relationship with a victim of trauma creates a vulnerable situation for the medical professional to develop vicarious trauma(Aparicio et al., 2013), with some researchers viewing this as inevitable, natural, and normal (Sansbury et al., 2015). In a qualitative study looking at the impact of vicarious trauma on trauma researchers, the results indicated that simply listening to traumatic stories (even in the context of research) \ldots & (health care professionals) AND (vicarious trauma); (health care professionals) AND (vicarious trauma); (health care professionals) AND (vicarious trauma); (health care professionals) AND (vicarious trauma); (health care professionals) AND (vicarious trauma); (health care professionals) AND (vicarious trauma) \\
6 & Physician Burnout: Assessing and addressing its impact on your family  “A physician struggling under an unbearable weight of emotions, or going to work feeling like they are enduring emotional trauma can have trickle down emotional effects on family members,” writes physician and US Navy veteran Caitlin Delaney, MD.  Like many industries, health care was catapulted into a new reality during the COVID-19 pandemic. But even as communities came together to express gratitude for caregivers risking \ldots & (health care professionals) AND (vicarious trauma) \\
7 & It is not uncommon that caregivers think last of their own well-being, leaving them vulnerable, frequently overworked and overwhelmed.   As research demonstrates, caregivers in good physical, emotional and mental health have a greater capacity to support their clients and are better able to perform their duties. However, continuous exposure to the trauma of others can further lead caregivers to experience symptoms of the individuals they serve. Researchers found that between 40\% and 85\% of \ldots & (health care professionals) AND (vicarious trauma) \\
8 & 3 MIN READ | Mental Health  Dennis Relojo-Howell  Why Mental Health Matters for Healthcare Professionals  Cite This Dennis Relojo-Howell, (2022, March 11). Why Mental Health Matters for Healthcare Professionals. Psychreg on Mental Health. https://www.psychreg.org/mental-health-matters-healthcare-professionals/ Reading Time: 3 minutes  Mental health is often viewed as a taboo topic, but it is something that we should be talking about more openly. Mental health affects everyone, and it’s \ldots & (health care professionals) AND (vicarious trauma) \\
9 & Helping Hurts  by Andy Conrad, MA, PLPC, NCC   Those serving in helping professions share a common feeling– that soul-deep responsibility for those under their care. Then, that empathic nature leaves many wondering if more could have been done, imagining alternate scenarios in the vacuum of hindsight, and questioning professional self-worth. These burdensome cycles of thought are products of the already distressing experiences of witnessing the physical and emotional contortions of trauma. \ldots & (healthcare workers) AND (vicarious trauma) \\
10 & Practice Self-Care to Be a Better Caregiver  Share on facebook Share on twitter Share on linkedin  All of us are caregivers in one form or another. We care for children, spouses, partners, parents, siblings, pets, colleagues, friends and our community. It may be because of a personal relationship, our career path or because we volunteer. We do it to help nurture well-being. Caregivers find personal satisfaction in giving of themselves because, in some way, it provides validation, connection, \ldots & (healthcare workers) AND (vicarious trauma) \\
11 & From powerlessness to recognition the meaning of palliative care clinicians’ experience of suffering  ABSTRACT Palliative care (PC) clinicians work alongside people who are at the end of their lives. These patients face death and suffering, which may also cause significant suffering for the PC clinicians themselves. Previous studies suggest that a significant number of PC professionals suffer from compassion fatigue,vicarious trauma and burnout. However, very few studies have attempted to \ldots & (health care professionals) AND (vicarious trauma) \\
12 & Identifying and Addressing Physician Burnout Before Impacting Your Family  Today is Physician Family Day, the perfect time to reflect and address any potential burnout symptoms and spend quality time re-engaging with your loved ones.  Like many industries, health care was catapulted into a new reality during the COVID-19 pandemic. But even as communities came together to express gratitude for caregivers risking their lives to treat patients amid the pandemic, another crisis was looming in the \ldots & (health care professionals) AND (vicarious trauma) \\
13 &     The most affected appeared to be drawn to the images, watching them over and over again and "re-traumatising themselves".    But vicarious trauma- known to affect healthcare workers and court jurors - can occur as a result of indirect exposure to other people's suffering.  Dr Pam Ramsden, from the University of Bradford, who led the new research, said: "Social media has enabled violent stories and graphic images to be watched by the public in unedited horrific detail.     Dr Ramsden, who \ldots & (healthcare workers) AND (vicarious trauma) \\
14 & Psychological impact of COVID‐19 pandemic on health care professionals and workers  The outbreak of a novel coronavirus starting from December 2019 and reaching pandemic proportions has raised concerns as to the ability of the current protective measures and the health‐care system to handle such a threat. Health‐care workers may experience considerable psychological distress as a result of the coronavirus disease‐19 pandemic due to providing direct patient care,vicarious trauma, quarantine, or \ldots & (health care professionals) AND (vicarious trauma) \\
15 & It is not uncommon that caregivers think last of their own well-being, leaving them vulnerable, frequently overworked and overwhelmed.    As research demonstrates, caregivers in good physical, emotional and mental health have a greater capacity to support their clients and are better able to perform their duties. However, continuous exposure to the trauma of others can further lead caregivers to experience symptoms of the individuals they serve. Researchers found that between 40\% and 85\% of \ldots & (health care professionals) AND (vicarious trauma) \\
16 & The Relationship Between Secondary Trauma and Distress  Understanding the impact of secondary trauma and mitigating risk with trauma-informed leadership.  Secondary trauma is not uncommon in the healthcare industry and can seriously impact the health of providers, patients, and organizations. In fact, one-in-four nurses suffers from post-traumatic stress disorder (PTSD) during their career. A recent study found that in critical care units, up to 48\% of nurses meet diagnostic criteria for the \ldots & (healthcare providers) AND (secondary traumatic stress) \\
17 & The Relationship Between Secondary Trauma and Distress   Secondary trauma is not uncommon in the healthcare industry and can seriously impact the health of providers, patients, and organizations. In fact, one-in-four nurses suffers from post-traumatic stress disorder (PTSD) during their career. A recent study found that in critical care units, up to 48\% of nurses meet diagnostic criteria for the disorder. No healthcare worker is immune to the risks of workplace trauma, especially in intensive \ldots & (healthcare providers) AND (secondary traumatic stress) \\
18 & Cost of Caring: How empathy and compassion affect health care workers  Rutgers University  Rutgers expert explains how health care workers can be affected by empathy and compassion  Compassion fatigue is just one of many consequences of working through the pandemic as a health care worker, according to Shari Munch, an associate professor at Rutgers School of Social Work.  “Dealing with constant suffering and trauma, on top of the double-edged sword of empathy and compassion, the cost of caring \ldots & (health care professionals) AND (compassion fatigue); (health care professionals) AND (compassion fatigue); (health care professionals) AND (compassion fatigue); (health care professionals) AND (compassion fatigue); (health care professionals) AND (compassion fatigue) \\
19 & Belfast Telegraph  UK Website Of The Year  Social media users 'traumatised'   Scientists investigated the "vicarious" psychological effect of images relating to distressing events such as the 9/11 Twin Tower terror attacks, the murder of fusilier Lee Rigby, and the disappearance of three-year-old Madeleine McCann.  They found that 22\% of 189 randomly chosen study participants were suffering symptoms commonly associated with Post Traumatic Stress Disorder (PTSD) triggered by what they had seen \ldots & (healthcare workers) AND (vicarious trauma) \\
20 & NewsCrime World  Graphic material on social media leading to Post Traumatic Stress Disorder symptoms  Scientists investigated the "vicarious" psychological effect of images relating to distressing events such as the 9/11terror attacks  Large numbers of people are being traumatised by violent and disturbing photos of news events on social media, research suggests.  Scientists investigated the "vicarious" psychological effect of images relating to distressing events such as the 9/11 Twin Tower \ldots & (healthcare workers) AND (vicarious trauma) \\
\end{longtable}

\small
\begin{longtable}{>{\raggedright\arraybackslash}p{0.06\linewidth}>{\raggedright\arraybackslash}p{0.68\linewidth}>{\raggedright\arraybackslash}p{0.20\linewidth}}
\caption{Top 20 results for the claim People watching graphic news reports can experience vicarious trauma} \\
\toprule
Rank & text & from\_queries \\
\midrule
\endfirsthead
\toprule
Rank & text & from\_queries \\
\midrule
\endhead
\midrule
\endfoot
\bottomrule
\endlastfoot
1 & What is Vicarious Trauma? What is Secondary Traumatic Stress?  Adapted from “The Compassion Fatigue Workbook  What is Vicarious Trauma PDF?   Vicarious trauma (VT) and Secondary Traumatic Stress (STS) are frequently used interchangeably to refer to the indirect trauma that can occur when we are exposed to difficult or disturbing images and stories second-hand.  This can occur by viewinggraphic news reports, gruesome or frightening television shows and various other media, hearing a detailed \ldots & graphic news reports \\
2 & Defining Vicarious Trauma and Secondary Traumatic Stress  What is Vicarious Trauma? What is Secondary Traumatic Stress?  Adapted from “The Compassion Fatigue Workbook  What is Vicarious Trauma PDF?   Vicarious trauma (VT) and Secondary Traumatic Stress (STS) are frequently used interchangeably to refer to the indirect trauma that can occur when we are exposed to difficult or disturbing images and stories second-hand.  This can occur by viewinggraphic news reports, gruesome or frightening \ldots & graphic news reports \\
3 & Vicarious trauma and the conviction of an Edinburgh student  Content Warning: mentions sexual assault, trauma.  Felix Beck, a fourth year University of Edinburgh student, has been convicted of sexually assaulting a first year student by the High Court. In a climate where fewer than a third of young men prosecuted for rape are convicted, any conviction feels like something worth celebrating.  Beck is yet to be sentenced and has been told by the Judge “you should not assume you will retain your \ldots & secondhand trauma \\
4 & 10 Common Things that Mentally Strong People Do During a Pandemic  Real-life cases and research findings show how you can handle and reduce your anxiety and unnecessary stress during COVID-19.  Photo by Heather Ford on Unsplash  Most of us know the basics: it is vital to create daily habits, to exercise, and to connect with our social supports using virtual meetings and social media. The COVID-19 global pandemic, extended lock-down and the systemic ramifications are not only unprecedented but \ldots & (vicarious trauma) AND (news exposure) \\
5 & Tracy S. Hutchinson, Ph.D.  The Pulse of Mental Health  10 Ways Mentally Strong People Handle a Pandemic  Research reveals how you can reduce your anxiety and stress during COVID-19.  Posted Apr 19, 2020  The COVID-19 global pandemic and the systemic ramifications are not only unprecedented but harrowing. We are all adjusting to new realities and grieving the loss of old ones. All of us are simply trying to manage our daily lives taking care of our children or parents, dealing with financial \ldots & (vicarious trauma) AND (news exposure) \\
6 & 10 Things Mentally Strong People Do During a Pandemic  Research reveals how you can reduce your anxiety and stress during COVID-19.  Image Credit: Unsplash  The COVID-19 global pandemic and the systemic ramifications are not only unprecedented but harrowing. We are all adjusting to new realities and grieving the loss of old ones. All of us are simply trying to manage our daily lives taking care of our children or parents, dealing with financial stresses, and adjusting to a new way to live.  All \ldots & (vicarious trauma) AND (news exposure) \\
7 & Skip to main content  Verified by Psychology Today  10 Ways to Stay Mentally Strong During a Pandemic  Research reveals how you can reduce your anxiety and stress during COVID-19.  The COVID-19 global pandemic and the systemic ramifications are not only unprecedented but harrowing. We are all adjusting to new realities and grieving the loss of old ones. All of us are simply trying to manage our daily lives—taking care of our children or parents, dealing with financial stresses, and adjusting to \ldots & (vicarious trauma) AND (news exposure) \\
8 & 10 Things Mentally Strong People Do During a Pandemic  Research reveals how you can reduce your anxiety and stress during COVID-19.  Image Credit: Unsplash  The COVID-19 global pandemic and the systemic ramifications are not only unprecedented but harrowing. We are all adjusting to new realities and grieving the loss of old ones. All of us are simply trying to manage our daily lives taking care of our children or parents, dealing with financial stresses, and adjusting to a new way to live.  All \ldots & (vicarious trauma) AND (news exposure) \\
9 & cing compassion fatigue, which is a common problem among people who work with those suffering from trauma.   Compassion fatigue can be career-ending, and it can also affect a nurse’s health and well-being. So, what is compassion fatigue in nursing, and what’s the best way to prevent or treat it? Here’s what you need to know.   What is Compassion Fatigue?   Also known as secondhand traumatization or “vicarious traumatization,” compassion fatigue is a state of emotional distress that results from \ldots & secondhand trauma \\
10 & 10 Ways to Stay Mentally Strong During a Pandemic  Research reveals how you can reduce your anxiety and stress during COVID-19.  Posted Apr 19, 2020 | Reviewed by Kaja Perina  The COVID-19 global pandemic and the systemic ramifications are not only unprecedented but harrowing. We are all adjusting to new realities and grieving the loss of old ones. All of us are simply trying to manage our daily lives—taking care of our children or parents, dealing with financial stresses, and adjusting to a \ldots & (vicarious trauma) AND (news exposure) \\
11 & 10 Ways Mentally Strong People Handle a Pandemic  Research reveals how you can reduce your anxiety and stress during COVID-19.  Posted Apr 19, 2020  The COVID-19 global pandemic and the systemic ramifications are not only unprecedented but harrowing. We are all adjusting to new realities and grieving the loss of old ones. All of us are simply trying to manage our daily lives taking care of our children or parents, dealing with financial stresses, and adjusting to a new way to live.  All \ldots & (vicarious trauma) AND (news exposure) \\
12 & NewsU.S. and the WorldNewsy   Anti-Asian hate taking mental toll on AAPI community  Stop Asian Hate Chicago Rally Posted at 1:02 PM, May 14, 2021 and last updated 2021-05-14 15:06:58-04  The toll of anti-Asian incidents during this pandemic goes beyond physical attacks. For many Asian Americans and Pacific Islanders— or AAPIs — it could also lead to sleep troubles, depression and PTSD.  That's because, psychologists we spoke to say anti-Asian hate impacts not only the victims — it can also have \ldots & doom scrolling \\
13 & 10 Things Mentally Strong People Do During A Pandemic  10 Things Mentally Strong People Do During A Pandemic   Research reveals how you can reduce your anxiety and stress during COVID-19 The COVID-19 global pandemic and the systemic ramifications are not only unprecedented but harrowing. We are all adjusting to new realities and grieving the loss of old ones. All of us are simply trying to manage our daily lives taking care of our children, parents, dealing with financial stresses, and a new \ldots & (vicarious trauma) AND (news exposure) \\
14 & 10 Things Mentally Strong People Do During a Pandemic   Image Credit: Unsplash  The COVID-19 global pandemic and the systemic ramifications are not only unprecedented but harrowing. We are all adjusting to new realities and grieving the loss of old ones. All of us are simply trying to manage our daily lives taking care of our children or parents, dealing with financial stresses, and adjusting to a new way to live.  All stressors related to COVID-19 may likely result in a myriad of negative \ldots & (vicarious trauma) AND (news exposure) \\
15 & themighty logo  10 Things I'm Doing to Help Overcome My Trauma  Hi Mighty readers. I wanted to start by sharing a few facts that are not always apparent to many:  Trauma can be big or small; this doesn’t dictate the level of pain it will have for someone.  Trauma can hurt even when it is not full-blown post-traumatic stress disorder (PTSD) or complex PTSD (C-PTSD).  Trauma may affect one person, but not another.  Trauma can manifest physically, in the body, as well as the mind.  Trauma can \ldots & secondhand trauma \\
16 & 10 Things Mentally Strong People Do During A Pandemic  10 Things Mentally Strong People Do During A Pandemic   Most of us know the basics: it is vital to create daily habits, to exercise, and to connect with our social supports using virtual meetings and social media. However, here are some suggestions based on evidence that may help you not only manage the pandemic but decrease the likelihood of long term mental health issues as a result of the coronavirus pandemic.  1. They limit news and \ldots & (vicarious trauma) AND (news exposure) \\
17 & 10 Things Mentally Strong People Do During a Pandemic  Research reveals how you can reduce your anxiety and stress during COVID-19.  The COVID-19 global pandemic and the systemic ramifications are not only unprecedented but harrowing. We are all adjusting to new realities and grieving the loss of old ones. All of us are simply trying to manage our daily lives taking care of our children or parents, dealing with financial stresses, and adjusting to a new way to live.  All stressors related to \ldots & (vicarious trauma) AND (news exposure) \\
18 & Israel Trauma and Media Project: Provides self-help tools and support to Israeli and Palestinian media professionals, whose coverage is impacted by burnout andsecondhand traumaassociated with job-stress and the violence to which they are exposed. Through trainings, educational activities and joint meetings over an 18 month period in cooperation with Israeli and Palestinian partners, media professionals in this region develop awareness of the role that trauma plays in exacerbating conflicts and \ldots & secondhand trauma \\
19 & Doom scrolling — the act of absorbing large amounts of online content, facilitated by the endless scroll ability of many social media platforms — is a distinctly modern phenomenon. It relies on the endless availability of content online, specifically on sites like Instagram, Facebook, and Twitter. The reasons people are prone to doom scrolling are nothing new, and as always, some people are more drawn to it than others.  Though it might feel like a harmless time-waster, doom scrolling has real \ldots & graphic media content \\
20 & director of the movie Total Recall, said, "I think it's a kind of purifying experience to see violence."  Violent computer games are also considered cathartic just like violent movies and dramas. Killing enemies and being deeply involved in the game environment is thought to be a useful outlet for the "primal human urge to kill". An imaginary founder, Dr Bartha, in promotional materials for fictional CyberDivision movement says, "We kill. It's OK. It's not our fault any more than breathing or \ldots & emotional desensitization \\
\end{longtable}

\small
\begin{longtable}{>{\raggedright\arraybackslash}p{0.06\linewidth}>{\raggedright\arraybackslash}p{0.68\linewidth}>{\raggedright\arraybackslash}p{0.20\linewidth}}
\caption{Top 20 results for the claim Symptoms of vicarious trauma can include anxiet} \\
\toprule
Rank & text & from\_queries \\
\midrule
\endfirsthead
\toprule
Rank & text & from\_queries \\
\midrule
\endhead
\midrule
\endfoot
\bottomrule
\endlastfoot
1 & Remember the last time you watched a really good movie? Who was the character you related to most, and how did you feel as they faced their ups and downs in the movie? That ability to put yourself in their shoes is a powerful human instinct we often call empathy. It helps us come together and relate to one another as human beings, but it can have a cost. One situation where empathy can cost us is when we are exposed to traumatic events that someone else has experienced. When we empathise with \ldots & (vicarious trauma) AND (hypervigilance) \\
2 & Sunday, November 28, 2010  Vicarious Trauma  The subject of vicarious trauma has been on my mind lately.  It's always on the back burner anyway, because of my profession. Health care workers, social workers, clergy members, humanitarian aid providers, counselors, and other individuals who confront human suffering on a regular basis are at risk for it. Unlike countertransference (the redirection of a service provider's feelings or unresolved conflicts toward the person being served) and \ldots & (vicarious trauma) AND (hypervigilance) \\
3 & Treating Trauma Self-care for providers Bookmark and Share  Indirect Trauma  Professionals who work with trauma survivors in an open, engaged, and empathic way and who feel responsible or committed to helping them are likely to experience indirect trauma. That means that they will be transformed by the work. The way helpers understand and experience the world and themselves is changed as they enter into the world of the survivor. While trauma work can be very meaningful and rewarding, it can \ldots & empathic strain \\
4 &   By Brandon Vogel   It is well known that lawyers are exposed to the trauma that their clients experience and disproportionately suffer from high levels of depression and substance use compared to social workers, mental health professionals and doctors.  These were among several key takeaways from Dr. Kerry O’Hara, Psy.D. (DBT Wellness \& Psychological Services) on the recent CLE Webinar, “Vicarious Trauma is Real.” She discussed the concept, also known as secondary trauma or compassion \ldots & (vicarious trauma) AND (hypervigilance) \\
5 &   Helping the World's Heroes        Reveal, Release, Reconnect  Follow Second-Hand Shock on Facebook    VTI's newest book, is available in our online store.   Book Reviews ViveBene Online Bookstore   Second-Hand Shock™  Second-Hand Shock orVicarious Traumais a silent thief that has the potential to rob the caring listener of their health and well-being.  Second-Hand Shock...what image does that conjure up for you? We’ve all heard of second-hand smoke. Put yourself in a room with a heavy smoker. \ldots & (vicarious trauma) AND (warning signs) \\
6 & cing compassion fatigue, which is a common problem among people who work with those suffering from trauma.   Compassion fatigue can be career-ending, and it can also affect a nurse’s health and well-being. So, what is compassion fatigue in nursing, and what’s the best way to prevent or treat it? Here’s what you need to know.   What is Compassion Fatigue?   Also known as secondhand traumatization or “vicarious traumatization,” compassion fatigue is a state of emotional distress that results from \ldots & secondhand trauma \\
7 & Put the Focus on Victims skip navigation   Key Terms  Trauma- and Healing-Related Terms  Acute stress disorder: Symptoms include flashbacks, a weakened immune system, fatigue, anxiety, hypersensitivity, generalized fear, and depression. The symptoms last from a minimum of 2 days to a maximum of 4 weeks following a traumatic event.  Affect: A pattern of observable behaviors that express victims' emotions. Common examples of affect are sadness, elation, and anger. Affect is different from mood \ldots & (vicarious trauma) AND (hypervigilance) \\
8 & Quick Answer: What Are The Warning Signs OfCompassion Fatigue?  How can nurses cope with compassion fatigue?  Over time, it can have a profound impact on a nurse’s professional and personal life, due to continual exposure to direct or secondary traumatic events.  Make time for yourself through self-care.  Regularly do things that you enjoy (e.g.  gardening, exercising, journaling).  Manage your family-work balance..  What is secondary traumatic stress disorder?  Secondary traumatic stress is \ldots & (compassion fatigue) AND (caregiver stress) \\
9 & compassion fatigue vs vicarious trauma  I refused to minimise the awfulness of their experience by including myself within their number. Instead, let’s talk about how it really feels to face the darkest corners of human life (and death). If anyone spends a lot of time with those who are suffering, an important question is how much suffering can be borne.  Define vicarious trauma, compassion fatigue, countertransference and distinguish from burnout Identify vicarious trauma impacts Describe the \ldots & (vicarious trauma) AND (warning signs); (vicarious trauma) AND (warning signs) \\
10 & On any given day, a firefighter, paramedic or police officer can see more violence and trauma than the average civilian may encounter in their entire lifetime.  The result of which, research suggests, may lead to damaging psychological effects. This is something we don't always consider when we see our friends and family who serve in these jobs.  Moreover, there is a large contingent of volunteers who serve our communities in these roles. Many people sign up to make a difference and serve their \ldots & (vicarious trauma) AND (hypervigilance) \\
11 & Vicarious trauma and the conviction of an Edinburgh student  Content Warning: mentions sexual assault, trauma.  Felix Beck, a fourth year University of Edinburgh student, has been convicted of sexually assaulting a first year student by the High Court. In a climate where fewer than a third of young men prosecuted for rape are convicted, any conviction feels like something worth celebrating.  Beck is yet to be sentenced and has been told by the Judge “you should not assume you will retain your \ldots & secondhand trauma \\
12 & Healing Body Memory of Trauma     Why do People Sexually Harass?      Science of Trust        Trauma and Restoration  When you experience a negative memory, do you experience an inability to move or take action? These “stuck” or “frozen” states are indicative of trauma. The trauma may be related to a single overwhelming event and/or it may be from a developmental disturbance, like childhood abuse or neglect. Trauma is about powerlessness, not being able to DO something helpful within the \ldots & (vicarious trauma) AND (hypervigilance) \\
13 & Put the Focus on Victims skip navigation   Ease Recovery  Tailoring Your Services  Having a victim-centered approach requires you to look comprehensively at victims' needs and recommend how various agencies in your SART could best provide for each need in a coordinated fashion. Such an approach might mean that you minimize the number of victim interviews, assist with restitution, provide safe places for victims to wait when they are attending court hearings, inform them about victim rights, \ldots & (vicarious trauma) AND (hypervigilance) \\
14 & Compassion Fatigue Kitchen Table 5 part series  In this training series, compassion fatigue is presented through five one-hour workshops that allow a more flexible, experiential introduction to the topic. These workshops involve lectures, exercises, worksheets and take-home activities.  Hour One: Understanding Compassion Fatigue andVicarious Trauma  Learning Objectives:    1. Defining compassion fatigue and vicarious trauma   2. Identifying compassion fatigue   3. Contrasting primary vs. \ldots & (vicarious trauma) AND (warning signs) \\
15 & Kitchen Table Groups on Compassion Fatigue  In this training series, compassion fatigue is presented through five one-hour workshops that allow a more flexible, experiential introduction to the topic. These workshops involve lectures, exercises, worksheets and take-home activities.  Hour One: Understanding Compassion Fatigue andVicarious Trauma  Learning Objectives:    1. Defining compassion fatigue and vicarious trauma   2. Identifying compassion fatigue   3. Contrasting primary vs. secondary \ldots & (vicarious trauma) AND (warning signs) \\
16 & WHEN: January 29, 2016  Workshop-Compassion Fatigue: Heal Thyself (3-CEUs)   Compassion Fatigue: Heal Thyself explores how vicarious stress and trauma can impact a clinician’s emotional health. Consideration is given to the importance of being mindful of one’s personal wellbeing. If we are compromised with stress and experience “burn-out,” we potentially place our clients at risk of receiving less than quality support. Focus will be placed on identifying signs of secondary trauma and self-care \ldots & (compassion fatigue) AND (self-care strategies); (compassion fatigue) AND (self-care strategies) \\
17 & PTSD Among Family Caregivers  We most often hear the acronym PTSD as it relates to those returning from the stresses of battle in a time of war.  As a caregiver for an older adult, you may have recognized some similarities in your own emotional response to your situation.  They are similar in many ways.  While your own life might not be in danger (though in many ways it could), the life of someone for whom you care is very much in the balance.  When trying to balance the various emotional, \ldots & (compassion fatigue) AND (caregiver stress); (compassion fatigue) AND (caregiver stress) \\
18 & We most often hear the acronym PTSD as it relates to those returning from the stresses of battle in a time of war.  As a caregiver for an older adult, you may have recognized some similarities in your own emotional response to your situation.  They are similar in many ways.  While your own life might not be in danger (though in many ways it could), the life of someone for whom you care is very much in the balance.  When trying to balance the various emotional, medical and financial aspects of a \ldots & (compassion fatigue) AND (caregiver stress); (compassion fatigue) AND (caregiver stress) \\
19 & Health Take-Away | Heads up on concussions: Separating myths from the truths   Understanding the diagnosis and treatment of head concussions, one of the most common traumas affecting millions of people each year, particularly during the fall sports season, requires separating some persistent myths from the essential truths.  Let's start with the truths:  TRUTH: Concussions are far more common — and unreported — than many people realize. An estimated 1.6 to 3.8 million sports- and \ldots & cognitive disruption \\
20 & Chat with us, powered by LiveChat  Caregiver PTSD  Caregiver PTSD2020-01-20T22:27:28-07:00  Caregivers suffering from PTSD is a real and serious problem. The most common manifestations of caregiver stress are compassion fatigue and burnout spawned by chronic feelings of anger, sadness, anxiety and guilt.  However, if a caregiver is exposed to a jarring traumatic experience, such as witnessing a loved one die suddenly of a heart attack or seeing them struggle with an excruciating ailment, they \ldots & (compassion fatigue) AND (caregiver stress) \\
\end{longtable}

\small
\begin{longtable}{>{\raggedright\arraybackslash}p{0.06\linewidth}>{\raggedright\arraybackslash}p{0.68\linewidth}>{\raggedright\arraybackslash}p{0.20\linewidth}}
\caption{Top 20 results for the claim Symptoms of vicarious trauma can include burnout} \\
\toprule
Rank & text & from\_queries \\
\midrule
\endfirsthead
\toprule
Rank & text & from\_queries \\
\midrule
\endhead
\midrule
\endfoot
\bottomrule
\endlastfoot
1 & Resources for Advocate Resilience  While advocating for victims of crime is rewarding and meaningful work, many advocates experience challenges related to job stress and indirect trauma exposure. Working with people who have experienced trauma and injustice can be overwhelming. Over half of all victim advocates experience severe traumatic stress symptoms and high levels of compassion fatigue(Benuto et al., 2018; Conrad and Kellar-Guenther, 2006) and 1-in-6 meet the PTSD diagnostic criteria from \ldots & (compassion fatigue) AND (emotional exhaustion) \\
2 & Resources for Advocate Resilience  While advocating for victims of crime is rewarding and meaningful work, many advocates experience challenges related to job stress and trauma exposure. Working with people who have experienced trauma and injustice can be overwhelming. Over half of all victim advocates experience severe traumatic stress symptoms and high levels of compassion fatigue(et al., 2018; Conrad and Kellar-Guenther, 2006) and 1-in-6 meet the PTSD diagnostic criteria from vicarious \ldots & (compassion fatigue) AND (depersonalization) \\
3 & The Cost of Caring:  Course Home Lesson 1 Lesson 2 Lesson 3 Lesson 4 Message Board  Lesson 2: Post-Traumatic Stress Disorders and Secondary Trauma  back 1 2 3 4 5 6 next  Secondary Traumatic Stress, Burnout and Vicarious Trauma  Secondary traumatic stress is a risk we incur when we engage empathically with an adult or child who has been traumatized. Charles Figley (1995) defines secondary traumatic stress as "the natural consequent behaviors resulting from knowledge about a traumatizing event \ldots & (secondary traumatic stress) AND (emotional exhaustion) \\
4 & Compassionate Fatigue Coaching  Compassionate Fatigue, Vicarious Trauma and Burn Out are complimentary, yet different from one another.  Compassionate Fatigue is an occupational hazard it is characterised as being the gradual profound emotional and physical exhaustion that takes place in workers over time, a gradual erosion of empathy, hope and compassion.  Vicarious Trauma is being constantly exposed to others trauma /distress and is the transformation that takes place, that changes your \ldots & (compassion fatigue) AND (emotional exhaustion) \\
5 & Stress and Burnout: A Hidden Occupational Hazard  by Kristen Paquette January 31, 2011  Image of Kristen Paquette  We all throw around terms like “stress” and “burnout” as though they should be regular parts of everyday life. We might say, “Oh man, am I stressed out!” Or, “Did you see Tom during today’s staff meeting? He totally snapped! He is really burnt out.”  Sure, we all have stress at work. In fact, some stress is healthy for us. But we need to be careful not to ignore the dangers of \ldots & (compassion fatigue) AND (emotional exhaustion) \\
6 & incredible-marketing Arrow  How Mental Health Practitioners Are Dealing with Trauma and Other Mental Illnesses During COVID-19   Frontline workers during the COVID-19 pandemic have been dealing with a pandemic of their own: mental illnesses like anxiety, depression, insomnia, distress, and trauma. Mental health practitioners are dealing with their own set of challenges during this time, and need to help themselves in order to help others. It is important to recognize and explore the mental \ldots & (compassion fatigue) AND (depersonalization) \\
7 &   Helping the World's Heroes        Reveal, Release, Reconnect  Follow Second-Hand Shock on Facebook    VTI's newest book, is available in our online store.   Book Reviews ViveBene Online Bookstore   Second-Hand Shock™  Second-Hand Shock orVicarious Traumais a silent thief that has the potential to rob the caring listener of their health and well-being.  Second-Hand Shock...what image does that conjure up for you? We’ve all heard of second-hand smoke. Put yourself in a room with a heavy smoker. \ldots & (vicarious trauma) AND (warning signs) \\
8 & Quick Answer: What Are The Warning Signs OfCompassion Fatigue?  How can nurses cope with compassion fatigue?  Over time, it can have a profound impact on a nurse’s professional and personal life, due to continual exposure to direct or secondary traumatic events.  Make time for yourself through self-care.  Regularly do things that you enjoy (e.g.  gardening, exercising, journaling).  Manage your family-work balance..  What is secondary traumatic stress disorder?  Secondary traumatic stress is \ldots & (compassion fatigue) AND (caregiver stress) \\
9 & , the caregiver will likely experience burnout.  It should be noted that burnout can be harmful to the individual receiving personal care and to the caregiver’s own health and well-being.  What are the Signs of Caregiver Burnout?  Experiencing one or more of the following symptoms may indicate that a caregiver is at risk of burning out.  1. Anxiety  When a caregiver experiences overwhelming worry thoughts, usually about the future and other everyday concerns such as finances.  2. Anger  Feeling \ldots & (compassion fatigue) AND (caregiver stress) \\
10 & WHEN: January 29, 2016  Workshop-Compassion Fatigue: Heal Thyself (3-CEUs)   Compassion Fatigue: Heal Thyself explores how vicarious stress and trauma can impact a clinician’s emotional health. Consideration is given to the importance of being mindful of one’s personal wellbeing. If we are compromised with stress and experience “burn-out,” we potentially place our clients at risk of receiving less than quality support. Focus will be placed on identifying signs of secondary trauma and self-care \ldots & (compassion fatigue) AND (self-care strategies); (compassion fatigue) AND (self-care strategies) \\
11 & compassion fatigue vs vicarious trauma  I refused to minimise the awfulness of their experience by including myself within their number. Instead, let’s talk about how it really feels to face the darkest corners of human life (and death). If anyone spends a lot of time with those who are suffering, an important question is how much suffering can be borne.  Define vicarious trauma, compassion fatigue, countertransference and distinguish from burnout Identify vicarious trauma impacts Describe the \ldots & (vicarious trauma) AND (warning signs); (vicarious trauma) AND (warning signs) \\
12 & Click here to see the meta data of this asset.  What isCompassion Fatigue?  Why talk about compassion fatigue when we are talking about childhood trauma?  Working with children and their families is tremendously rewarding, but engaging with the chronic traumas children experience in their families can be stressful, overwhelming and engender hopelessness. Given the knowledge and training gap most physicians have about addressing developmental trauma, as well as the high stakes and complexity of \ldots & (compassion fatigue) AND (emotional exhaustion); (secondary traumatic stress) AND (emotional exhaustion) \\
13 & Kitchen Table Groups on Compassion Fatigue  In this training series, compassion fatigue is presented through five one-hour workshops that allow a more flexible, experiential introduction to the topic. These workshops involve lectures, exercises, worksheets and take-home activities.  Hour One: Understanding Compassion Fatigue andVicarious Trauma  Learning Objectives:    1. Defining compassion fatigue and vicarious trauma   2. Identifying compassion fatigue   3. Contrasting primary vs. secondary \ldots & (vicarious trauma) AND (warning signs) \\
14 & Compassion Fatigue Kitchen Table 5 part series  In this training series, compassion fatigue is presented through five one-hour workshops that allow a more flexible, experiential introduction to the topic. These workshops involve lectures, exercises, worksheets and take-home activities.  Hour One: Understanding Compassion Fatigue andVicarious Trauma  Learning Objectives:    1. Defining compassion fatigue and vicarious trauma   2. Identifying compassion fatigue   3. Contrasting primary vs. \ldots & (vicarious trauma) AND (warning signs) \\
15 & Trauma Centre Cameroon Peter Essoka Trauma Centre Cameroon \&quot;I love my work but lately I find it contaminating my personal life. I have nightmares about the horrible things I hear about from clients, my sex life has deteriorated, I'm irritable and distractible, I'm afraid for my kids and tend to overprotect them, and I don't trust anybody anymore. I don't know what is happening to me.\&quot; A Therapist Double-Edged Sword “Empathetic engagement helps workers meet the needs of traumatized \ldots & (secondary traumatization) AND (burnout) \\
16 & The Cost of Caring: 10 Ways to Prevent Compassion Fatigue   Written March 26, 2020  Compassion fatigue can be a serious occupational hazard for those in any kind of helping profession, with a majority of those in the field reporting experiencing at least some degree of it in their lives. This is no surprise, as it is typically those with the most empathy who are the most at risk.  Compassion fatigue is characterized by physical and emotional exhaustion and a profound decrease in the ability to \ldots & (secondary traumatic stress) AND (emotional exhaustion) \\
17 & Treating Trauma Self-care for providers Bookmark and Share  Indirect Trauma  Professionals who work with trauma survivors in an open, engaged, and empathic way and who feel responsible or committed to helping them are likely to experience indirect trauma. That means that they will be transformed by the work. The way helpers understand and experience the world and themselves is changed as they enter into the world of the survivor. While trauma work can be very meaningful and rewarding, it can \ldots & empathic strain \\
18 & Tag Archives: vicarious trauma  Physician heal thyself…  And nurses, counselors, therapists, caregivers, social workers, case managers, community health workers, teachers, first responders, peer outreach… in short any person who spends their time (work or volunteer) opening themselves to the experiences of others’ suffering. In a recent continuing education exercise, I was asked to examine myself for resilience and potential risk of compassion fatigue. The simple act of participating in the \ldots & secondary traumatization \\
19 & Chat with us, powered by LiveChat  Caregiver PTSD  Caregiver PTSD2020-01-20T22:27:28-07:00  Caregivers suffering from PTSD is a real and serious problem. The most common manifestations of caregiver stress are compassion fatigue and burnout spawned by chronic feelings of anger, sadness, anxiety and guilt.  However, if a caregiver is exposed to a jarring traumatic experience, such as witnessing a loved one die suddenly of a heart attack or seeing them struggle with an excruciating ailment, they \ldots & (compassion fatigue) AND (caregiver stress) \\
20 & Share this  Top Tips From Nurses on Dealing With Burnout  by Timon Kaple • 5 min read Top Tips From Nurses on Dealing With Burnout  Experts report that compassion fatigue, often referred to colloquially as "burnout," can take a physical and mental toll on nurses, first responders, physicians, and other professionals who provide patient care or work in emergency situations. According to the Substance Abuse and Mental Health Services Administration (SAMHSA), compassion fatigue consists of two \ldots & (secondary traumatic stress) AND (emotional exhaustion) \\
\end{longtable}

\small
\begin{longtable}{>{\raggedright\arraybackslash}p{0.06\linewidth}>{\raggedright\arraybackslash}p{0.68\linewidth}>{\raggedright\arraybackslash}p{0.20\linewidth}}
\caption{Top 20 results for the claim Symptoms of vicarious trauma can include changes in outlook on life} \\
\toprule
Rank & text & from\_queries \\
\midrule
\endfirsthead
\toprule
Rank & text & from\_queries \\
\midrule
\endhead
\midrule
\endfoot
\bottomrule
\endlastfoot
1 & Vicarious TraumaAmong Helpers  Vicarious Trauma Management and Self-Care  An article by Secure Start Intern, Tara Hearne  Vicarious trauma is a term used to describe the reaction of those who work in a helping role in response to hearing their clients’ trauma stories. Also known as secondary traumatisation or compassion fatigue,vicarious trauma changes a helper’s view about themselves, others and the world via alteration of cognitive schemas of identity, memory systems and belief systems. \ldots & (vicarious trauma) AND (cognitive schemas) \\
2 & Vicarious TraumaAmong Helpers  Vicarious Trauma Management and Self-Care An article by Secure Start Intern, Tara Hearne Vicarious trauma is a term used to describe the reaction of those who work in a helping role in response to hearing their clients’ trauma stories. Also known as secondary traumatisation or compassion fatigue,vicarious trauma changes a helper’s view about themselves, others and the world via alteration of cognitive schemas of identity, memory systems and belief systems. \ldots & (vicarious trauma) AND (cognitive schemas) \\
3 & What isVicarious Trauma?  Vicarious trauma is a term used to describe the potential mental health impacts for workers who are repeatedly responding to the trauma of others. This can happen in many types of workplaces, but particularly where the core business of the organisation is to offer support to clients who have directly suffered the impacts of trauma. Examples of these types of services include child protection, family violence, corrections, youth justice, and housing services, amongst \ldots & (vicarious trauma) AND (cognitive schemas) \\
4 & Dissertation/Thesis Abstract  The Relationship Between Vicarious Trauma and Sexual Desire Among Behavioral Health Clinicians by Branson, Dana C., Ph.D., Northcentral University, 2011, 166; 3458587 Abstract (Summary)  Vicarious trauma refers to the psychological changes that can occur in behavioral health clinicians (BHC) as a result of joining with clients in an empathic relationship. Vicarious trauma is a significant occupational hazard for BHCs, with potentially negative effects on the \ldots & constructivist self-development theory \\
5 & Treating Trauma Self-care for providers Bookmark and Share  Indirect Trauma  Professionals who work with trauma survivors in an open, engaged, and empathic way and who feel responsible or committed to helping them are likely to experience indirect trauma. That means that they will be transformed by the work. The way helpers understand and experience the world and themselves is changed as they enter into the world of the survivor. While trauma work can be very meaningful and rewarding, it can \ldots & empathic strain \\
6 & compassion fatigue vs vicarious trauma  I refused to minimise the awfulness of their experience by including myself within their number. Instead, let’s talk about how it really feels to face the darkest corners of human life (and death). If anyone spends a lot of time with those who are suffering, an important question is how much suffering can be borne.  Define vicarious trauma, compassion fatigue, countertransference and distinguish from burnout Identify vicarious trauma impacts Describe the \ldots & (vicarious trauma) AND (warning signs); (vicarious trauma) AND (warning signs) \\
7 &   Helping the World's Heroes        Reveal, Release, Reconnect  Follow Second-Hand Shock on Facebook    VTI's newest book, is available in our online store.   Book Reviews ViveBene Online Bookstore   Second-Hand Shock™  Second-Hand Shock orVicarious Traumais a silent thief that has the potential to rob the caring listener of their health and well-being.  Second-Hand Shock...what image does that conjure up for you? We’ve all heard of second-hand smoke. Put yourself in a room with a heavy smoker. \ldots & (vicarious trauma) AND (warning signs) \\
8 & Date of Conferral        Maria van Tilburg   Vicarious trauma can impact anyone working with a traumatized person. Theconstructivist self-development theoryasserts that vicarious trauma can negatively distort how the helper thinks about the world and can cause increased stress. Researchers have explored stress and coping models and have studied how increased stress can negatively impact coping and health behaviors. However, researchers have not explored how vicarious trauma, coping, and health \ldots & constructivist self-development theory \\
9 & Date of Conferral        Maria van Tilburg   Vicarious trauma can impact anyone working with a traumatized person. Theconstructivist self-development theoryasserts that vicarious trauma can negatively distort how the helper thinks about the world and can cause increased stress. Researchers have explored stress and coping models and have studied how increased stress can negatively impact coping and health behaviors. However, researchers have not explored how vicarious trauma, coping, and health \ldots & constructivist self-development theory \\
10 & Vicarious TraumaAmong Sexual Assault Nurse Examiners  ABSTRACT Vicarious trauma (VT), the phenomenon of changes in cognition and worldview that result from empathic response and repeated exposure to narratives of trauma, is a risk for helping professionals. This descriptive, correlational study sought to examine levels of VT among sexual assault nurse examiners (SANEs) as compared with other women’s health nurses. It also explored whether levels of VT are different for nurses who have \ldots & (vicarious trauma) AND (worldview) \\
11 &   This research examined how contact with the legal, medical, and mental health systems affects rape survivors' psychological well-being. Although community services may be beneficial for some victims, there is increasing evidence that they can add trauma, rather than alleviate distress (termed secondary victimization). This study examined how secondary victimization affects rape survivors' posttraumatic stress (PTS) symptoms. Adaptive and snowball sampling were used to recruit a sample of 102 \ldots & constructivist self-development theory \\
12 & When Compassion Hurts: An Autoethnographic Experience ofVicarious Traumaby a Forensic Nurse  Caring for victims of violence includes obtaining history, examining and collecting evidentiary matter, and providing expert testimony in the court of law. In this paper, the researcher shared her own experiences as a forensic nurse on how being compassionate and listening to the stories of victims of violence and witnessing the sustained injuries led to some emotional stress which is referred to in the \ldots & (vicarious trauma) AND (worldview); (vicarious trauma) AND (worldview) \\
13 & Surviving Sexual Assault  Surviving Sexual Assault  Surviving Sexual Assault  By Hung Tran, Psy.D.  Sexual assault is any sexual act, which is unwelcome or unwanted. It is sometimes committed through use of manipulation, coercion, intimidation, threats, force, or a controlled substance. It can range from sexual battery, to threat of sexual assault, to rape.  For survivors of sexual assault, talking about your experience may by particularly painful and difficult. For every survivor of sexual \ldots & (vicarious trauma) AND (reactions) \\
14 & Vicarious traumatization in healthcare providers in response to COVID-19 pandemic in Kelantan, Malaysia  Background In the fight against the COVID-19 pandemic, frontline healthcare providers who are engaged in the direct diagnosis, treatment, and care of patients face a high risk of infection yet receive inadequate protection from contamination and minimal support to cope with overwork, frustration, and exhaustion. These problems have created significant psychological and mental health concerns \ldots & (secondary traumatic stress) AND (cognitive schemas) \\
15 & Your Account Vicarious Trauma and Disaster Mental Health Understanding Risks and Promoting Resilience (To see other currencies, click on price) Add to basket     Main description:  Vicarious Trauma and Disaster Mental Health focuses on the clinician and the impact of working with disaster survivors. Floods, hurricanes, tornadoes, mass shootings, terrorism and other large-scale catastrophic events have increased in the last decade and disaster resilience has become a national imperative. This \ldots & (secondary traumatic stress) AND (cognitive schemas) \\
16 & Compassion Fatigue Kitchen Table 5 part series  In this training series, compassion fatigue is presented through five one-hour workshops that allow a more flexible, experiential introduction to the topic. These workshops involve lectures, exercises, worksheets and take-home activities.  Hour One: Understanding Compassion Fatigue andVicarious Trauma  Learning Objectives:    1. Defining compassion fatigue and vicarious trauma   2. Identifying compassion fatigue   3. Contrasting primary vs. \ldots & (compassion fatigue) AND (warning signs); (vicarious trauma) AND (warning signs) \\
17 & that levels of VPTG was higher than ambulance officers in Australia, which is due to socio-cultural influences. This can be explained through China’s collectivist culture, with workers working with more pride and responsibility, which reduces distress and promotes the process of growth” (Kang et al., 2018).  Australian Paramedic: Higher levels of Vicarious Post-Traumatic Growth due to socio-cultural influences  Different workplaces has shown to have an influence on the levels of Vicarious \ldots & constructivist self-development theory \\
18 & Kitchen Table Groups on Compassion Fatigue  In this training series, compassion fatigue is presented through five one-hour workshops that allow a more flexible, experiential introduction to the topic. These workshops involve lectures, exercises, worksheets and take-home activities.  Hour One: Understanding Compassion Fatigue andVicarious Trauma  Learning Objectives:    1. Defining compassion fatigue and vicarious trauma   2. Identifying compassion fatigue   3. Contrasting primary vs. secondary \ldots & (compassion fatigue) AND (warning signs); (vicarious trauma) AND (warning signs) \\
19 & The Experiences of Male Counselors of Children Who Have Experienced Trauma  Kathleen Michelle Wallace, Walden University   Men are increasingly underrepresented in counselor education and in the counseling profession, with only 27\% of members of the American Counseling Association reporting as men. Men in counseling often feel marginalized and isolated. Additionally, they are socialized to be independent, emotionally and physically strong, and to focus on success, while being discouraged from \ldots & constructivist self-development theory \\
20 & Vicarious Traumaand Disaster Mental Health: Understanding Risks and Promoting Resilience Gertie Quitangon Editor Mark R. Evces Editor mixed material bibliography New York Routledge Taylor \& Francis Group 2015 en English xviii, 207 pages; 23 x 16 cm. Vicarious Trauma and Disaster Mental Health focuses on the clinician and the impact of working with disaster survivors. Floods, hurricanes, tornadoes, mass shootings, terrorism and other large-scale catastrophic events have increased in the last \ldots & (vicarious trauma) AND (cognitive schemas); (secondary traumatic stress) AND (cognitive schemas) \\
\end{longtable}

\small
\begin{longtable}{>{\raggedright\arraybackslash}p{0.06\linewidth}>{\raggedright\arraybackslash}p{0.68\linewidth}>{\raggedright\arraybackslash}p{0.20\linewidth}}
\caption{Top 20 results for the claim Symptoms of vicarious trauma can include depression} \\
\toprule
Rank & text & from\_queries \\
\midrule
\endfirsthead
\toprule
Rank & text & from\_queries \\
\midrule
\endhead
\midrule
\endfoot
\bottomrule
\endlastfoot
1 & Compassionate Fatigue Coaching  Compassionate Fatigue, Vicarious Trauma and Burn Out are complimentary, yet different from one another.  Compassionate Fatigue is an occupational hazard it is characterised as being the gradual profound emotional and physical exhaustion that takes place in workers over time, a gradual erosion of empathy, hope and compassion.  Vicarious Trauma is being constantly exposed to others trauma /distress and is the transformation that takes place, that changes your \ldots & (compassion fatigue) AND (emotional exhaustion) \\
2 & autious.   2. Trust needs – an individual’s capacity to trust that their emotional, physical and psychological needs will be met by others, as well as the ability to trust their own perceptions and beliefs. Shifts in cognitive schemas of basic trust present as suspiciousness of others, even in once trustworthy relationships, and stifled self-trust leading to self-doubt.   3. Esteem needs – an individual’s inherent value for self and others. Changes in cognitive schemas of esteem needs can cause \ldots & (vicarious trauma) AND (symptomatology) \\
3 & cing compassion fatigue, which is a common problem among people who work with those suffering from trauma.   Compassion fatigue can be career-ending, and it can also affect a nurse’s health and well-being. So, what is compassion fatigue in nursing, and what’s the best way to prevent or treat it? Here’s what you need to know.   What is Compassion Fatigue?   Also known as secondhand traumatization or “vicarious traumatization,” compassion fatigue is a state of emotional distress that results from \ldots & secondhand trauma \\
4 & 10 Ways to Improve Resilience \& PreventDepression\& Anxiety by Hacking the Kynurenine Pathway  I wrote this in response to someone who works with survivors of domestic violence and wants to incorporate more exercise into their life. None of this should be taken as medical advice. Consult your primary care provider before changing anything about your healthcare.   How Stress Causes Neuropsychiatric Disorders:  First, I must say I admire the work you’re doing. All work is hard, but helping the \ldots & (vicarious trauma) AND (depression) \\
5 & , physical and psychological needs will be met by others, as well as the ability to trust their own perceptions and beliefs. Shifts in cognitive schemas of basic trust present as suspiciousness of others, even in once trustworthy relationships, and stifled self-trust leading to self-doubt.   3. Esteem needs – an individual’s inherent value for self and others. Changes in cognitive schemas of esteem needs can cause feelings of inadequacy, doubt in own abilities and reduced esteem for others as \ldots & (vicarious trauma) AND (symptomatology); (vicarious trauma) AND (symptomatology) \\
6 & Resources for Advocate Resilience  While advocating for victims of crime is rewarding and meaningful work, many advocates experience challenges related to job stress and indirect trauma exposure. Working with people who have experienced trauma and injustice can be overwhelming. Over half of all victim advocates experience severe traumatic stress symptoms and high levels of compassion fatigue(Benuto et al., 2018; Conrad and Kellar-Guenther, 2006) and 1-in-6 meet the PTSD diagnostic criteria from \ldots & (compassion fatigue) AND (emotional exhaustion) \\
7 & Stress and Burnout: A Hidden Occupational Hazard  by Kristen Paquette January 31, 2011  Image of Kristen Paquette  We all throw around terms like “stress” and “burnout” as though they should be regular parts of everyday life. We might say, “Oh man, am I stressed out!” Or, “Did you see Tom during today’s staff meeting? He totally snapped! He is really burnt out.”  Sure, we all have stress at work. In fact, some stress is healthy for us. But we need to be careful not to ignore the dangers of \ldots & (compassion fatigue) AND (emotional exhaustion) \\
8 & Treating Trauma Self-care for providers Bookmark and Share  Indirect Trauma  Professionals who work with trauma survivors in an open, engaged, and empathic way and who feel responsible or committed to helping them are likely to experience indirect trauma. That means that they will be transformed by the work. The way helpers understand and experience the world and themselves is changed as they enter into the world of the survivor. While trauma work can be very meaningful and rewarding, it can \ldots & empathic strain \\
9 &   Helping the World's Heroes        Reveal, Release, Reconnect  Follow Second-Hand Shock on Facebook    VTI's newest book, is available in our online store.   Book Reviews ViveBene Online Bookstore   Second-Hand Shock™  Second-Hand Shock orVicarious Traumais a silent thief that has the potential to rob the caring listener of their health and well-being.  Second-Hand Shock...what image does that conjure up for you? We’ve all heard of second-hand smoke. Put yourself in a room with a heavy smoker. \ldots & (vicarious trauma) AND (warning signs) \\
10 & tel: 07722761483  PIE - Supporting people through trauma.  “Anyone in the helping professions is at risk of compassion fatigue, physical and emotional exhaustion, and a reduced ability to empathise. I found a shocking lack of awareness at senior management level about vicarious trauma and its impact upon staff well-being” Anon The Guardian 2015 "The cost of caring."  It’s not only service users who experience emotional distress which can lead to feelings of helplessness and frustration but also \ldots & (compassion fatigue) AND (emotional exhaustion) \\
11 & compassion fatigue vs vicarious trauma  I refused to minimise the awfulness of their experience by including myself within their number. Instead, let’s talk about how it really feels to face the darkest corners of human life (and death). If anyone spends a lot of time with those who are suffering, an important question is how much suffering can be borne.  Define vicarious trauma, compassion fatigue, countertransference and distinguish from burnout Identify vicarious trauma impacts Describe the \ldots & (vicarious trauma) AND (warning signs); (vicarious trauma) AND (warning signs) \\
12 & Resources >   Psychological Trauma Article  Psychological Trauma – What Every Trauma Worker Should Know  Zoe Lodrick MSc. BA Hons (1st). Dip (psych). Dip (couns). UKCP    In this paper I outline the basis of psychological trauma.  I surmise what happens to the human brain and nervous system at the point of trauma, and during trauma recall (both choiceful and intrusive).  I discuss how, and why, some people may be more susceptible to developing traumatic symptomatology, and possible reasons for \ldots & (vicarious trauma) AND (symptomatology) \\
13 & 4. Rethinking trauma in parenting traumatised children  As stated above, a traumatic incident is defined as an ‘emotional shock following a stressful event’ and ‘results from events outside normal human experience which overwhelm usual coping mechanisms’. Traumatic incidents, and traumatisation, can be subdivided into primary trauma, secondary trauma,vicarious trauma and trans-generational trauma. People susceptible to primary trauma are usually present when the traumatic incident occurs; \ldots & (vicarious trauma) AND (symptomatology) \\
14 & Compassion Fatigue Kitchen Table 5 part series  In this training series, compassion fatigue is presented through five one-hour workshops that allow a more flexible, experiential introduction to the topic. These workshops involve lectures, exercises, worksheets and take-home activities.  Hour One: Understanding Compassion Fatigue andVicarious Trauma  Learning Objectives:    1. Defining compassion fatigue and vicarious trauma   2. Identifying compassion fatigue   3. Contrasting primary vs. \ldots & (vicarious trauma) AND (warning signs) \\
15 & Resources >   Psychological Trauma Article  Psychological Trauma – What Every Trauma Worker Should Know     In this paper I outline the basis of psychological trauma.  I surmise what happens to the human brain and nervous system at the point of trauma, and during trauma recall (both choiceful and intrusive).  I discuss how, and why, some people may be more susceptible to developing traumatic symptomatology, and possible reasons for the repeat victimization experienced by so many \ldots & (vicarious trauma) AND (symptomatology) \\
16 & Vicarious trauma and the conviction of an Edinburgh student  Content Warning: mentions sexual assault, trauma.  Felix Beck, a fourth year University of Edinburgh student, has been convicted of sexually assaulting a first year student by the High Court. In a climate where fewer than a third of young men prosecuted for rape are convicted, any conviction feels like something worth celebrating.  Beck is yet to be sentenced and has been told by the Judge “you should not assume you will retain your \ldots & secondhand trauma \\
17 & Treatment and prevention professionals often experience vicarious and secondary trauma that can lead to burnout and negatively impact services to clients. Participants will be given the opportunity to reflect upon how this reality affects their personal and professional lives and identify personal and organizational practices that can help mitigate the risks and improve services.  Participants will engage in a series of activities that are designed to help them assess the impact that vicarious \ldots & trauma stewardship \\
18 & Kitchen Table Groups on Compassion Fatigue  In this training series, compassion fatigue is presented through five one-hour workshops that allow a more flexible, experiential introduction to the topic. These workshops involve lectures, exercises, worksheets and take-home activities.  Hour One: Understanding Compassion Fatigue andVicarious Trauma  Learning Objectives:    1. Defining compassion fatigue and vicarious trauma   2. Identifying compassion fatigue   3. Contrasting primary vs. secondary \ldots & (vicarious trauma) AND (warning signs) \\
19 & Skip to main content  Self-efficacy and humanitarian aid workers   As a result of frequent exposure to trauma, aid workers are at high risk for negative psychological symptoms. Training specifically geared at fostering critical incident self-efficacy in humanitarian aid workers may bolster critical incident self-efficacy as well as general self-efficacy as they relate to experiences of traumatic symptomatology and resilience. Sixty-three aid workers completed questionnaires regarding efficacy, \ldots & (vicarious trauma) AND (symptomatology) \\
20 & Short report: social identification, quality of professional life and emotional maladjustment in healthcare providers during the first wave of the COVID-19 emergency.  This research analyzed the effect of professional, organizational and care-unit identifications on both healthcare professionals' quality of professional life and mental health. This research was done in a local hospital in a region of northern Italy which was one of the first regions to be impacted by the first wave of the \ldots & (vicarious trauma) AND (depression) \\
\end{longtable}

\small
\begin{longtable}{>{\raggedright\arraybackslash}p{0.06\linewidth}>{\raggedright\arraybackslash}p{0.68\linewidth}>{\raggedright\arraybackslash}p{0.20\linewidth}}
\caption{Top 20 results for the claim Therapists can experience vicarious trauma} \\
\toprule
Rank & text & from\_queries \\
\midrule
\endfirsthead
\toprule
Rank & text & from\_queries \\
\midrule
\endhead
\midrule
\endfoot
\bottomrule
\endlastfoot
1 & 18 Minutes  Edited \& clinically reviewed by THE BALANCE Team Fact checked  Vicarious trauma, often termed secondary trauma is defined as indirectly experiencing a traumatic incident through a first-hand account or description of the situation. Therapists and counselors as well as emergency crews, law enforcement officers, physicians, and judges, and lawyers may be vulnerable to vicarious trauma. Any individual who has a close connection with a trauma victim may be subjected to secondary \ldots & (counselors) AND (vicarious trauma) \\
2 & "Women Working Together" in the Shadows of Sexual Violence  The impact of trauma work upon practitioners is well documented. Experiences of vicarious trauma, compassion fatigue are explored throughout the trauma literature. However, less understood are the emotional benefits of such work and the restorative nature of working in trauma settings.   Introduction A limited number of empirical studies have considered the opposite of vicarious trauma, known as vicarious resilience which is the \ldots & (practitioners) AND (trauma work) \\
3 & webinar register page  Understanding vicarious trauma in the context of Covid 19 and racial injustice This program, when attended in its entirety offers 1.0 APA CE for Psychologists and 1.0 BBS California CEU forCounselors, Social Workers and Marriage and Family Therapists.  Mental Health professionals routinely provide care to individuals with significant trauma histories. The incessant responsiveness and empathic regard that is required when giving care to trauma clients can result in the \ldots & (counselors) AND (vicarious trauma); (counselors) AND (vicarious trauma); (counselors) AND (compassion fatigue) \\
4 & • Working with psychological trauma: an interpretative phenomenological analysis of trauma-focused CBT and EMDR        Folland, Caroline H; University of Derby (2017-04-12)       Purpose: The Diagnostic and Statistical Manual of Mental Disorders (DSM-5) criteria for post-traumatic stress disorder (PTSD), acknowledged repeated or extreme indirect exposure to aversive details of traumatic event(s) in the course of professional duties, can lead to symptoms of PTSD. This has led to discussions \ldots & (therapists) AND (vicarious trauma) \\
5 & Vicarious Trauma  Vicarious Trauma: Awareness, Balance, and Creating Connections Posted on 05/03/2017  Our community has experienced a lot of tragedies this past year. Through these tragedies, the power of “community” has truly been demonstrated by the compassion and help to those in need by everyone around. There have been teachers, principals, and mental health professionals that have provided assistance to those impacted by the tragedies. These events can create trauma for those directly \ldots & (mental health professionals) AND (vicarious trauma); (mental health professionals) AND (vicarious trauma); (mental health professionals) AND (vicarious trauma) \\
6 & Vicarious Trauma- May 3, 2017  vicarious trauma: awareness, balance, and creating connections. Firefighter upset or hurt with events that have unfolded  Our community has experienced a lot of tragedies this past year. Through these tragedies, the power of “community” has truly been demonstrated by the compassion and help to those in need by everyone around. There have been teachers, principals, and mental health professionals that have provided assistance to those impacted by the tragedies. \ldots & (mental health professionals) AND (vicarious trauma) \\
7 & Hidden Occupational Hazards for Social Service Providers  Burnout,secondary traumatic stress(STS), and vicarious trauma (VT) are occupational hazards for those who work in social service settings, specifically social workers, therapists,counselors, correctional rehabilitative staff, and case managers who encounter the physical, emotional, and/or sexual trauma of others. Research surrounding these phenomena continues to expand, shedding light on the development, symptomology, potential risk and \ldots & (counselors) AND (secondary traumatic stress) \\
8 & Medical professionals such as physicians, nurses,psychotherapists, and emergency workers who help traumatized patients may develop their own Post Traumatic Stress Disorder (PTSD) symptoms as an indirect response to their patient’s suffering. This phenomenon has been referred to as compassion fatigue, vicarious traumatization or secondary traumatic stress.   A survey showed that “86.9\% of emergency response personnel reported symptoms after exposure to highly distressing events with traumatized \ldots & (psychotherapists) AND (compassion fatigue) \\
9 & Somatic Psychology  Bridging the mind-body gap  Compassion Fatigue  Bodily symptoms of empathy  Medical professionals such as physicians, nurses,psychotherapists, and emergency workers who help traumatized patients may develop their own Post Traumatic Stress Disorder (PTSD) symptoms as an indirect response to their patient’s suffering. This phenomenon has been referred to as compassion fatigue, vicarious traumatization or secondary traumatic stress.   A survey showed that “86.9\% of emergency \ldots & (psychotherapists) AND (compassion fatigue) \\
10 & The double-edged-sword effect of empathy: The secondary traumatic stress and vicarious posttraumatic growth of psychological hotline counselors during the outbreak of COVID-19  The novel Coronavirus Disease 2019 (COVID-19) outbreak increases public mental stress and pandemic-related trauma. Timely and comprehensive online mental health services have been deployed across China. Hotline counselors can be exposed to the stress and trauma experienced by callers indirectly. They may experience \ldots & (counselors) AND (secondary traumatic stress) \\
11 & Self-Care forMental Health Professionals  In his book, Trauma Informed-Mindfulness With Teens, Sam Himelstein stresses the need for self-care for mental health professionals dealing with traumatised teens.   His final chapter is devoted to self-care and professional practice.  Dr. Cirecie West-Olatunji, counsellor educator,  also stresses the need for self-care for professionals working with people experiencing traumatic stress.  Her video presentation is available as part of the courses \ldots & (mental health professionals) AND (vicarious trauma) \\
12 & Date of Award   Degree Name  Doctor of Philosophy   Counselor Education and Counseling Psychology  First Advisor  Dr. Glinda Rawls  Second Advisor  Dr. Stephen Craig  Third Advisor  Dr. Patricia Reeves   Trauma, African American, counselor(s), secondary trauma, race-based trauma, qualitative research   Nearly 83\% of Americans are exposed to a traumatic event (Benjet et al., 2016). Given this, every counselor will more than likely work with trauma survivors (Trippany et al., 2004). Because of \ldots & (counselors) AND (secondary traumatic stress) \\
13 & Resources for Advocate Resilience  While advocating for victims of crime is rewarding and meaningful work, many advocates experience challenges related to job stress and indirect trauma exposure. Working with people who have experienced trauma and injustice can be overwhelming. Over half of all victim advocates experience severe traumatic stress symptoms and high levels of compassion fatigue (Benuto et al., 2018; Conrad and Kellar-Guenther, 2006) and 1-in-6 meet the PTSD diagnostic criteria \ldots & (therapists) AND (vicarious trauma); (therapists) AND (vicarious trauma) \\
14 & April 21st, 2020  Vicarious Trauma - Tools \& Tips for Essential Workers  Lauren Shechter, LCSW, SEP  As the COVID-19 pandemic continues, essential workers are still bravely risking their own safety to provide much needed services for others. In the last two months, some might have started to notice adverse changes in their mental and physical wellbeing even if they are otherwise “physically healthy.” The essential workers I have encountered in my personal and professional lives have described \ldots & (mental health professionals) AND (vicarious trauma) \\
15 & Secondary Trauma  Secondary trauma is the emotional spread of the effect of trauma symptoms as a result of close and extended contact with traumatized individuals. Generally, the traumatized person has close and extended contact with others, such as family members, friends, or mental health professionals. These individuals are at risk of experiencing psychological distress through direct exposure to the traumatized person. Symptoms of secondary trauma are similar to those of posttraumatic \ldots & (psychotherapists) AND (compassion fatigue) \\
16 & Africa: Why Social Workers Suffer From Depression and Need Help Too  Every 40 seconds, someone, somewhere in the world, dies by suicide. On World \#MentalHealth Day, @WHO is calling for 40 seconds of action to prevent suicide. 10 October 2019  The task of helping people cope with traumatic events in their lives in an environment free of stigma is one that mental health professionals shoulder selflessly.  Patients are required to be as transparent as possible as they narrate what troubles them in \ldots & (mental health professionals) AND (vicarious trauma) \\
17 & Dilip Balu   Vicarious Trauma | Supervision | Collaborative Care | Trauma Aware Schooling   Dilip is a Senior Clinical Social Worker and Team Leader of the Child Protection Counselling Service based in the Nepean Blue Mountains Local Health District. Dilip also practices privately as a trainer and clinical supervisor for individuals and groups. His clinical experience ranges across Child Protection, Out of Home Care, Child and Adolescent Mental Health and long-term psychotherapy. Dilip has a \ldots & (clinicians) AND (vicarious trauma) \\
18 & Coming to Terms with Stress  Crime and negative police interactions become the lenses through which we see the world. Often we forget that those lenses are attached to glasses that we can take off.  Wearing those glasses all the time and believing that's the only way to view the world is a response to trauma.  Overwhelmed with sadness, confusion, and suppressed rage, I began to see a therapist. I dreaded each weekly appointment, because it meant that I'd have to confront the things I was \ldots & (mental health professionals) AND (vicarious trauma) \\
19 & Your Account Vicarious Trauma and Disaster Mental Health Understanding Risks and Promoting Resilience (To see other currencies, click on price) Add to basket     Main description:  Vicarious Trauma and Disaster Mental Health focuses on the clinician and the impact of working with disaster survivors. Floods, hurricanes, tornadoes, mass shootings, terrorism and other large-scale catastrophic events have increased in the last decade and disaster resilience has become a national imperative. This \ldots & (clinicians) AND (vicarious traumatization) \\
20 & Vicarious Traumaand Disaster Mental Health: Understanding Risks and Promoting Resilience Gertie Quitangon Editor Mark R. Evces Editor mixed material bibliography New York Routledge Taylor \& Francis Group 2015 en English xviii, 207 pages; 23 x 16 cm. Vicarious Trauma and Disaster Mental Health focuses on the clinician and the impact of working with disaster survivors. Floods, hurricanes, tornadoes, mass shootings, terrorism and other large-scale catastrophic events have increased in the last \ldots & (clinicians) AND (vicarious trauma); (clinicians) AND (vicarious traumatization) \\
\end{longtable}

\small
\begin{longtable}{>{\raggedright\arraybackslash}p{0.06\linewidth}>{\raggedright\arraybackslash}p{0.68\linewidth}>{\raggedright\arraybackslash}p{0.20\linewidth}}
\caption{Top 20 results for the claim Vicarious trauma can lead to symptoms similar to firsthand trauma} \\
\toprule
Rank & text & from\_queries \\
\midrule
\endfirsthead
\toprule
Rank & text & from\_queries \\
\midrule
\endhead
\midrule
\endfoot
\bottomrule
\endlastfoot
1 & What is vicarious trauma and how to prevent it in EMS  Moderating exposure to secondary trauma and practicing self-care can help EMTs and paramedics avoid PTSD  By Lieutenant Brad Bouchillon, Contributor to In Public Safety  Anyone who works in the emergency medical services, fire service, law enforcement or another public safety field has heard of post-traumatic stress disorder. PTSD is often caused after experiencing a traumatic event or enduring many impactful events over time.  However, \ldots & (vicarious trauma) AND (direct trauma) \\
2 & What is Vicarious Trauma? What is Secondary Traumatic Stress?  Adapted from “The Compassion Fatigue Workbook  What is Vicarious Trauma PDF?   Vicarious trauma (VT) and Secondary Traumatic Stress (STS) are frequently used interchangeably to refer to the in direct trauma that can occur when we are exposed to difficult or disturbing images and stories second-hand.  This can occur by viewing graphic news reports, gruesome or frightening television shows and various other media, hearing a detailed \ldots & (indirect trauma) AND (direct trauma) \\
3 & Home clinician What isVicarious Traumaand How to Prevent It What isVicarious Traumaand How to Prevent It  What isVicarious Traumaand How to Prevent It  Start a public administration degree at American Military University.  By Lieutenant Brad Bouchillon, Statesboro (GA) Fire Department  Anyone who works in the fire service, emergency medical services, law enforcement, or another public safety field has heard of post-traumatic stress disorder (PTSD). PTSD is often caused after experiencing a \ldots & (vicarious trauma) AND (direct trauma) \\
4 & Skip to main content  Vicarious trauma in healthcare providers: How to recognize \& heal from secondary trauma  With the COVID-19 pandemic nearing its second anniversary, chances are good that you've heard of vicarious trauma at some point. That's because a higher number of healthcare workers have been experiencing it lately. Still, the term may not mean much to you unless you've been through it yourself or know someone who has.  If you work in the healthcare field, it's a good idea to \ldots & (vicarious trauma) AND (direct trauma); (vicarious trauma) AND (direct trauma); (vicarious trauma) AND (direct trauma); (vicarious trauma) AND (direct trauma); (vicarious trauma) AND (direct trauma) \\
5 & Secondary trauma | Education Support Secondary trauma - teachers and education staff  Secondary trauma   What is secondary trauma?  Trauma can be described as an event, or events, which result in psychological effects and shock. Secondary trauma can be experienced where, for example a student or colleague has experienced a traumatic event, and you feel the impact of their distress. It can be summed up as the cumulative effect on the educator, of dealing with traumatic events which have happened \ldots & (secondary traumatic stress) AND (psychological effects) \\
6 & Areas of Specialization   You may hear people talk about being "traumatized" or discuss a "traumatizing event". One of the important things to understand about trauma is that it's not the event itself but rather the mind's reaction to the event that results in someone being "traumatized" to a greater or lesser degree. Because of this, two people could experience the same event and react very differently. Whether your trauma has been caused by one or multiple events, and whether that/those \ldots & (vicarious trauma) AND (direct trauma) \\
7 & cing compassion fatigue, which is a common problem among people who work with those suffering from trauma.   Compassion fatigue can be career-ending, and it can also affect a nurse’s health and well-being. So, what is compassion fatigue in nursing, and what’s the best way to prevent or treat it? Here’s what you need to know.   What is Compassion Fatigue?   Also known as secondhand traumatization or “vicarious traumatization,” compassion fatigue is a state of emotional distress that results from \ldots & secondhand trauma \\
8 & autious.   2. Trust needs – an individual’s capacity to trust that their emotional, physical and psychological needs will be met by others, as well as the ability to trust their own perceptions and beliefs. Shifts in cognitive schemas of basic trust present as suspiciousness of others, even in once trustworthy relationships, and stifled self-trust leading to self-doubt.   3. Esteem needs – an individual’s inherent value for self and others. Changes in cognitive schemas of esteem needs can cause \ldots & (vicarious trauma) AND (symptomatology) \\
9 & Trauma \& Trauma Therapy  What is trauma?   Trauma refers to any experiences that are emotionally painful or distressing, that overwhelm a person's ability to cope. Often traumatic experiences are considered those that are outside of a normal human experience or a threat to life or safety. However trauma can also be more subtle and may be more frequent for some people. Some examples of trauma include witnessing or experiencing violence or assault as well as other life circumstances such as child \ldots & (vicarious trauma) AND (direct trauma) \\
10 & , physical and psychological needs will be met by others, as well as the ability to trust their own perceptions and beliefs. Shifts in cognitive schemas of basic trust present as suspiciousness of others, even in once trustworthy relationships, and stifled self-trust leading to self-doubt.   3. Esteem needs – an individual’s inherent value for self and others. Changes in cognitive schemas of esteem needs can cause feelings of inadequacy, doubt in own abilities and reduced esteem for others as \ldots & (vicarious trauma) AND (symptomatology); (vicarious trauma) AND (symptomatology) \\
11 & Social Workers And Mental Health Providers  1477 Words6 Pages  Social workers and mental health providers work in a field where trauma is prevalent. This means that people who work in these types of fields are more likely to experience what is called “secondary traumatization”, “vicarious traumatization” or “compassion fatigue”. These terms all essentially describe the negative impact of bonding with a trauma survivor and exposure to the survivor’s traumatic encounters (McCann \& Pearlman 1990). \ldots & firsthand trauma \\
12 & Resources >   Psychological Trauma Article  Psychological Trauma – What Every Trauma Worker Should Know     In this paper I outline the basis of psychological trauma.  I surmise what happens to the human brain and nervous system at the point of trauma, and during trauma recall (both choiceful and intrusive).  I discuss how, and why, some people may be more susceptible to developing traumatic symptomatology, and possible reasons for the repeat victimization experienced by so many \ldots & (symptomatology) AND (primary trauma) \\
13 & Psychological Trauma – What Every Trauma Worker Should Know  Zoe Lodrick MSc. BA Hons (1st). Dip (psych). Dip (couns). UKCP    In this paper I outline the basis of psychological trauma.  I surmise what happens to the human brain and nervous system at the point of trauma, and during trauma recall (both choiceful and intrusive).  I discuss how, and why, some people may be more susceptible to developing traumatic symptomatology, and possible reasons for the repeat victimization experienced by so \ldots & (symptomatology) AND (primary trauma) \\
14 & Resources >   Psychological Trauma Article  Psychological Trauma – What Every Trauma Worker Should Know     In this paper I outline the basis of psychological trauma.  I surmise what happens to the human brain and nervous system at the point of trauma, and during trauma recall (both choiceful and intrusive).  I discuss how, and why, some people may be more susceptible to developing traumatic symptomatology, and possible reasons for the repeat victimization experienced by so many \ldots & (symptomatology) AND (primary trauma); (vicarious trauma) AND (symptomatology) \\
15 & Resources >   Psychological Trauma Article  Psychological Trauma – What Every Trauma Worker Should Know     In this paper I outline the basis of psychological trauma.  I surmise what happens to the human brain and nervous system at the point of trauma, and during trauma recall (both choiceful and intrusive).  I discuss how, and why, some people may be more susceptible to developing traumatic symptomatology, and possible reasons for the repeat victimization experienced by so many \ldots & (symptomatology) AND (primary trauma) \\
16 & 4. Rethinking trauma in parenting traumatised children  As stated above, a traumatic incident is defined as an ‘emotional shock following a stressful event’ and ‘results from events outside normal human experience which overwhelm usual coping mechanisms’. Traumatic incidents, and traumatisation, can be subdivided into primary trauma, secondary trauma,vicarious trauma and trans-generational trauma. People susceptible to primary trauma are usually present when the traumatic incident occurs; \ldots & (symptomatology) AND (primary trauma); (vicarious trauma) AND (symptomatology) \\
17 & 4. Rethinking trauma in parenting traumatised children  As stated above, a traumatic incident is defined as an ‘emotional shock following a stressful event’ and ‘results from events outside normal human experience which overwhelm usual coping mechanisms’. Traumatic incidents, and traumatisation, can be subdivided into primary trauma, secondary trauma, vicarious trauma and trans-generational trauma. People susceptible to primary trauma are usually present when the traumatic incident occurs; \ldots & (symptomatology) AND (primary trauma) \\
18 & Vicarious trauma and the conviction of an Edinburgh student  Content Warning: mentions sexual assault, trauma.  Felix Beck, a fourth year University of Edinburgh student, has been convicted of sexually assaulting a first year student by the High Court. In a climate where fewer than a third of young men prosecuted for rape are convicted, any conviction feels like something worth celebrating.  Beck is yet to be sentenced and has been told by the Judge “you should not assume you will retain your \ldots & secondhand trauma \\
19 & Resources >   Psychological Trauma Article  Psychological Trauma – What Every Trauma Worker Should Know  Zoe Lodrick MSc. BA Hons (1st). Dip (psych). Dip (couns). UKCP    In this paper I outline the basis of psychological trauma.  I surmise what happens to the human brain and nervous system at the point of trauma, and during trauma recall (both choiceful and intrusive).  I discuss how, and why, some people may be more susceptible to developing traumatic symptomatology, and possible reasons for \ldots & (symptomatology) AND (primary trauma); (vicarious trauma) AND (symptomatology) \\
20 & Resources >   Psychological Trauma Article  Psychological Trauma – What Every Trauma Worker Should Know  Zoe Lodrick MSc. BA Hons (1st). Dip (psych). Dip (couns). UKCP    In this paper I outline the basis of psychological trauma.  I surmise what happens to the human brain and nervous system at the point of trauma, and during trauma recall (both choiceful and intrusive).  I discuss how, and why, some people may be more susceptible to developing traumatic symptomatology, and possible reasons for \ldots & (symptomatology) AND (primary trauma) \\
\end{longtable}

\small
\begin{longtable}{>{\raggedright\arraybackslash}p{0.06\linewidth}>{\raggedright\arraybackslash}p{0.68\linewidth}>{\raggedright\arraybackslash}p{0.20\linewidth}}
\caption{Top 20 results for the claim Vicarious trauma is not the same as experiencing trauma firsthand} \\
\toprule
Rank & text & from\_queries \\
\midrule
\endfirsthead
\toprule
Rank & text & from\_queries \\
\midrule
\endhead
\midrule
\endfoot
\bottomrule
\endlastfoot
1 & Home clinician What isVicarious Traumaand How to Prevent It What isVicarious Traumaand How to Prevent It  What isVicarious Traumaand How to Prevent It  Start a public administration degree at American Military University.  By Lieutenant Brad Bouchillon, Statesboro (GA) Fire Department  Anyone who works in the fire service, emergency medical services, law enforcement, or another public safety field has heard of post-traumatic stress disorder (PTSD). PTSD is often caused after experiencing a \ldots & (vicarious trauma) AND (empathic engagement) \\
2 & What isVicarious Trauma?  Vicarious trauma is a term used to describe the potential mental health impacts for workers who are repeatedly responding to the trauma of others. This can happen in many types of workplaces, but particularly where the core business of the organisation is to offer support to clients who have directly suffered the impacts of trauma. Examples of these types of services include child protection, family violence, corrections, youth justice, and housing services, amongst \ldots & (vicarious trauma) AND (empathic engagement) \\
3 &   The MSBA recently hosted a livestream presentation, Bearing Witness to Our Clients’ Trauma: Doing Our Jobs AND Taking Care of Ourselves. The program used clips from the movie “Just Mercy”, and featured Michael Kahn, ReelTime CLE Co-Founder, “Recovering Attorney”, Licensed Professional Counselor and Coach; Cathleen Price, renowned civil rights attorney with the Equal Justice Initiative and Adjunct Assistant Professor at Columbia University; and Sean O’Brien, Professor and clinical director at \ldots & (vicarious trauma) AND (empathic engagement) \\
4 & Vicarious trauma and the conviction of an Edinburgh student  Content Warning: mentions sexual assault, trauma.  Felix Beck, a fourth year University of Edinburgh student, has been convicted of sexually assaulting a first year student by the High Court. In a climate where fewer than a third of young men prosecuted for rape are convicted, any conviction feels like something worth celebrating.  Beck is yet to be sentenced and has been told by the Judge “you should not assume you will retain your \ldots & secondhand trauma \\
5 & Bounce Back  One of the greatest problems currently facing workers in the caring profession, specifically first responders, is the effect of vicarious trauma. Vicarious trauma is described as a transformation of a worker in a helping profession as a result of empathic engagement with trauma survivors and their traumatic material, combined with the responsibility and commitment to help them.  Bounce Back is an app to teach people how to manage and encourage further empathic responses. By \ldots & (vicarious trauma) AND (empathic engagement) \\
6 & cing compassion fatigue, which is a common problem among people who work with those suffering from trauma.   Compassion fatigue can be career-ending, and it can also affect a nurse’s health and well-being. So, what is compassion fatigue in nursing, and what’s the best way to prevent or treat it? Here’s what you need to know.   What is Compassion Fatigue?   Also known as secondhand traumatization or “vicarious traumatization,” compassion fatigue is a state of emotional distress that results from \ldots & secondhand trauma \\
7 & MitigatingVicarious Traumaand Countertransference Essay  Click here to ORDER an A++ paper from our MASTERS and DOCTORATE WRITERS: MitigatingVicarious Traumaand Countertransference Essay  MitigatingVicarious Traumaand Countertransference Essay     Carl Jung (1954) said, “You can exert no influence if you are not susceptible to influence.” When engaging with clients who have experienced trauma, practitioners are vulnerable to being impacted by the clients’ experiences. Exposure to multiple client \ldots & (vicarious trauma) AND (empathic engagement); (secondary traumatic stress) AND (empathic engagement) \\
8 & Managing Vicarious Trauma  Loading Events  « All Events    • This event has passed.  Managing Vicarious Trauma  27 November 2019  This one-day interactive professional development training explores the nature, dynamics and risks of vicarious trauma (VT), contrasts it with burnout, post-traumatic stress disorder and compassion fatigue, and supports you to stay healthy and safe in your work with people impacted by diverse traumas.  The training distinguishes the differences between, and \ldots & direct trauma exposure \\
9 & Compassion Fatigue  Michelle Hamilton, Compassion fatigue, registered psychologist  Compassion Fatigue (CF) is a term used to describe the potential emotional, physical, spiritual exhaustion that helping professionals can experience through repeated exposure to the emotional pain of clients.  CF can impact a helper’s personal and/or professional lives with symptoms such as emotional distancing or numbness, difficulty concentrating, intrusive imagery, loss of hope, exhaustion and irritability \ldots & (compassion fatigue) AND (empathic engagement) \\
10 & Compassion Fatigue  Compassion Fatigue  Compassion Fatigue (CF) is a term used to describe the potential emotional, physical, spiritual exhaustion that helping professionals can experience through repeated exposure to the emotional pain of clients.  CF can impact a helper’s personal and/or professional lives with symptoms such as emotional distancing or numbness, difficulty concentrating, intrusive imagery, loss of hope, exhaustion and irritability (Figley, 1995).  Compassion fatigue has been \ldots & (compassion fatigue) AND (empathic engagement) \\
11 & PTSD is a mental health condition triggered by eitherexperiencing trauma firsthandor witnessing or learning of a terrifying event that happens to others.  Post-Traumatic Stress Disorder  How do you react to a frightening event? We all respond differently to these types of occurrences for many reasons. Certain events that cause significant trauma can be diagnosed as post-traumatic stress disorder (PTSD), which can disrupt your job, relationships, health, and everyday activities you usually \ldots & experiencing trauma firsthand \\
12 & Skip to Content  Deploying a Trauma-Informed Approach: Use the 4 R’s  By Dr. Isaiah Pickens, PhD, Founder/CEO of iOpening Enterprises; Partner, PresenceLearning  This article was originally published in District Administration.  Our nation’s teachers are as critical to young people and society as they’ve ever been. Teachers provide care that might be lacking in the home. They’re trained to guard against school violence, and more recently,  the threat of COVID-19. They’re working with students \ldots & experiencing trauma firsthand \\
13 & Saturday, February 2, 2013  The Effects of Trauma on Clients and Mental Health Counselors              It seems over the past decade there have been more acts of violence than in previous times.  This causes an increase in traumatic events that Americans are facing which in turn creates fears of the event reoccuring, or feelings of anxiety, depression over the event(s) (Bell, Friel, \& Stacher, 2007).  Traumas have various effects on survivors, first responders, and even those who come to lend \ldots & (vicarious trauma) AND (empathic engagement) \\
14 & themighty logo  10 Things I'm Doing to Help Overcome My Trauma  Hi Mighty readers. I wanted to start by sharing a few facts that are not always apparent to many:  Trauma can be big or small; this doesn’t dictate the level of pain it will have for someone.  Trauma can hurt even when it is not full-blown post-traumatic stress disorder (PTSD) or complex PTSD (C-PTSD).  Trauma may affect one person, but not another.  Trauma can manifest physically, in the body, as well as the mind.  Trauma can \ldots & secondhand trauma \\
15 & 01/16/2014 05:33 pm ET Updated Mar 18, 2014  Sex Trafficking: Beyond the Abstract Evil  rubberball via Getty Images Readers of my work know I've investigated the horrors of sex trafficking in places like Thailand and Cambodia. But besides personally meeting the survivors and warriors who work daily to combat this complex crime, I've also become a bit of an anti-slavery conference nut. I've attended conferences in Chittagong, Manila, L.A., D.C. and N.Y.C. Some were funded by well-established \ldots & (vicarious trauma) AND (empathic engagement) \\
16 & Why 1 in 4 nurses suffers from PTSD (and how to help them)  Read Advisory Board's take: The four factors that undermine nurse resilience  Nearly 30\% of nurses suffer from post-traumatic stress disorder (PTSD) during their careers, contributing to high rates of turnover in the profession, Emilie Le Beau Lucchesi reports for the New York Times.  What is PTSD?  According to the DSM-5, the reference for psychiatric diagnoses, PTSD can stem fromexperiencing trauma firsthandas well as from witnessing \ldots & experiencing trauma firsthand \\
17 & Deploying a trauma-informed approach: Use the 4 Rs  As school leaders and educators, being equipped to deal with heightened and varying degrees of behavioral and mental health challenges among our students is as essential as ensuring their academic achievement. By: | February 9, 2021 Dr. Isaiah Pickens, PhD, is the Founder/CEO of iOpening Enterprises and a Partner with PresenceLearning.   Our nation’s teachers are as critical to young people and society as they’ve ever been. Teachers provide \ldots & experiencing trauma firsthand \\
18 & Document Type   Degree Name  Master of Social Work (MSW)   Social Work   Lyle S. Hallman Faculty of Social Work  First Advisor  Eliana Suarez  Advisor Role   Second Advisor  Jennifer Root  Advisor Role  Second Reader   What are the experiences of racialized frontline workers? How do they experience vicarious racism? I interviewed 8 frontline workers who were involved in counseling racialized individuals. The narrative paradigm, constructivist self-development theory (CSDT), and critical race \ldots & (vicarious trauma) AND (empathic engagement) \\
19 & On Trauma and Interfaith Grace  I started a course called “Pastoral Care” this quarter, and had no idea what to expect. It is week 2, and it feels like week 9. We tackled “trauma” today, at 9 am on a Friday morning. I had no idea what was coming.  Trauma is defined using fancy words and vocabulary, images and concepts. Reading about it this week, I found myself vaguely identifying with signs of trauma, but knew that everyone would have some feelings within the framework. I read a story about a \ldots & (vicarious trauma) AND (empathic engagement) \\
20 & another sense, one was not present. One was not there, at least not as a conscious being. In being overwhelmed in the moment, one immediately took oneself out of the experience as an immediate reaction and survival mechanism. The traumatic experience remains unintegrated with one’s other life experiences, spinning in a tight circle of repetition.   The circle of repetition is split off from the person’s awareness and everyday life, remaining isolated—“sequestered” is the technical term for \ldots & (empathic distress) AND (empathic engagement) \\
\end{longtable}

\small
\begin{longtable}{>{\raggedright\arraybackslash}p{0.06\linewidth}>{\raggedright\arraybackslash}p{0.68\linewidth}>{\raggedright\arraybackslash}p{0.20\linewidth}}

\caption{Top 20 results for the claim Vicarious trauma refers to the psychological impact on an individual who is exposed to the traumatic experiences of others through empathic engagement}
\label{tab:last_rank_table}\\
\toprule
Rank & text & from\_queries \\
\midrule
\endfirsthead
\toprule
Rank & text & from\_queries \\
\midrule
\endhead
\midrule
\endfoot
\bottomrule
\endlastfoot
1 & Vicarious at risk of being negatively impacted by  Vicarious trauma is the cumulative emotional residue that acounsellor may experience when working with people who have suffered traumaticlife events (Vicarious Trauma, 2018)Vicarious traumais a negative reaction to trauma exposure,and includes a range of psychosocial symptoms.  Any person who workswith victims of trauma is at risk of being negatively impacted by the effectsof vicarious trauma. Factors that makecounsellors or volunteers \ldots & (vicarious trauma) AND (trauma exposure) \\
2 & Work-RelatedVicarious Traumaand PTSD: What to Know and Do  Written by Dr Caroline Marlow, Chartered Psychologist and Director of L\&M Consulting Ltd. It was published by the Law Society (29.03.2019) but adapted here for cross-industry consideration.  Vicarious Trauma and PTSD: What To Know  If you work in services such as; the emergency and medical services, social services, law and justice, immigration, journalism and the media/social media, humanitarian support, clinical psychology, and the \ldots & (vicarious trauma) AND (trauma exposure) \\
3 & Treating Trauma Self-care for providers Bookmark and Share  Indirect Trauma  Professionals who work with trauma survivors in an open, engaged, and empathic way and who feel responsible or committed to helping them are likely to experience indirect trauma. That means that they will be transformed by the work. The way helpers understand and experience the world and themselves is changed as they enter into the world of the survivor. While trauma work can be very meaningful and rewarding, it can \ldots & empathic strain \\
4 & What is Complex Trauma?  What is Trauma? Trauma refers to the psychological distress that follows a very stressful event in which a person directly or indirectly experiences threat to her/his safety. Examples of traumatic events are: being trapped in a burning building, being robbed at gun point; subjected to child abuse or neglect, victim of a sexual assault. It is normal to experience distress in the face of such traumatic events. In the aftermath, some people do not recover fully, and suffer \ldots & (psychological distress) AND (traumatic events) \\
5 & Skip to main content  Vicarious trauma in healthcare providers: How to recognize \& heal from secondary trauma  With the COVID-19 pandemic nearing its second anniversary, chances are good that you've heard of vicarious trauma at some point. That's because a higher number of healthcare workers have been experiencing it lately. Still, the term may not mean much to you unless you've been through it yourself or know someone who has.  If you work in the healthcare field, it's a good idea to \ldots & (vicarious trauma) AND (trauma exposure) \\
6 & ManagingVicarious Trauma  Loading Events  « All Events    • This event has passed.  ManagingVicarious Trauma  27 November 2019  This one-day interactive professional development training explores the nature, dynamics and risks of vicarious trauma(VT), contrasts it with burnout, post-traumatic stress disorder and compassion fatigue, and supports you to stay healthy and safe in your work with people impacted by diverse traumas.  The training distinguishes the differences between, and appropriate \ldots & (vicarious trauma) AND (trauma exposure); (vicarious trauma) AND (trauma exposure) \\
7 & Multiple Exposure: Increased Risk  Image courtesy of Dr. Leslie Anne Ross, UCLA  This diagram presents the wide range of factors that can lead to a professional experiencing greater rates of stress and distress which can, over time, contribute to someone developing compassion fatigue and related symptoms.   Personal Vulnerabilities   The individual’s current life circumstance, their personal history, coping style and personality characteristics all affect how they are impacted by the work. In \ldots & (vicarious trauma) AND (trauma exposure) \\
8 & Presented by Zoe Lodrick  Zoe’s enthusiastic presentation style enabled her to provided a depth of knowledge on the impact of interpersonal psychological trauma. This knowledge initially related to the effect on the young person’s brain development and how this may, in later life, impact negatively on their current and future relationships (including social and work relationships) and life chances.  Zoe’s presentation was liberally peppered with examples of the negative experiences in the \ldots & (traumatic experiences) AND (therapeutic relationship) \\
9 & Forms of Trauma  An individual person’s experience of trauma and the impact it has on them varies from person to person. A person can also experience multiple forms trauma. For example, a young person who is growing up in a community in which they may be experiencing racism and may also be exposed to ongoing violence in their neighborhood or family.    Improving awareness of trauma  The first step to healing trauma is increasing our awareness of the various forms of trauma that people \ldots & empathic strain \\
10 & © 2018 by The Secondary Traumatic Stress Consortium.     • Facebook Social Icon   • Twitter Social Icon  Multiple Exposure: Increased Risk  Image courtesy of Dr. Leslie Anne Ross, UCLA  This diagram presents the wide range of factors that can lead to a professional experiencing greater rates of stress and distress which can, over time, contribute to someone developing compassion fatigue and related symptoms.   Personal Vulnerabilities   The individual’s current life circumstance, their personal \ldots & (vicarious trauma) AND (trauma exposure) \\
11 & humanization, intimidation, et cetera that are experienced by members of groups targeted by racism, heterosexism, ageism, ableism, sexism, and other forms of oppression, and groups impacted by poverty. Maria Root, who coined the term insidious trauma described the concepts as follows: “Traumatogenic effects of oppression that are not necessarily overtly violent or threatening to bodily well-being at the given moment but that do violence to the soul and spirit. ” (Root 1992; Brown \& Ballou, \ldots & (vicarious trauma) AND (trauma exposure) \\
12 & Post-Traumatic Stress Disorder \& Seafaring  Traumatic events are very common in any society and people can be traumatized by any event they perceive as life threatening. Traumatic events such as family and social violence, accidents, war, natural disasters and assaults of any nature can traumatize people diminishing their sense of safety leaving them temporarily and in some cases incapacitated to cope with the event ( Schwarz, L., 2016). Individuals suffering major traumatic events may \ldots & (psychological distress) AND (traumatic events) \\
13 & givers, this is everybody, I believe vicarious trauma, something that happens throughout the community, especially the more and more I look at the definition, cumulative effects of contact with survivors, right? So because I’m, I’m black, and I’m in the black community, I know more and more people who’ve been affected by this kind of like racial based violence or police brutality or poverty, or, you know, because I’m in New York, I know more people who’ve been affected by 911. Right, as but as \ldots & (vicarious trauma) AND (trauma exposure) \\
14 & Over one in ten of us suffered at least four traumatic experiences as children. So the likelihood of charities encountering trauma survivors among their beneficiaries without realising it is high indeed. Click To Tweet  At the most basic level, ‘trauma-informed approaches’ are ways of supporting people that recognise specific needs they may have as a result of past or ongoing trauma. Trauma can arise from many places, so it is difficult to estimate how many people in Britain are living with its \ldots & (psychological distress) AND (traumatic experiences) \\
15 & U.S. flag   Glossary of Terms  Burnout is a state of physical, emotional, and mental exhaustion caused by long-term involvement in emotionally demanding situations. Symptoms may include depression, cynicism, boredom, loss of compassion, and discouragement.  Compassion fatigue is a combination of physical, emotional, and spiritual depletion associated with caring for others who are in significant emotional pain and physical distress.  Compassion satisfaction refers to the pleasure derived from \ldots & (vicarious trauma) AND (trauma exposure) \\
16 & Mom Psych  Trauma and Resilience   Childhood Emotional Maltreatment Causes Troubled Romantic Relationships, Studies Suggest  Childhood Trauma Linked to Schizophrenia  APA Monitor: Treating Traumatized Children  Resilience: The Mental Muscle Everyone Has  The Road to Resilience  Building Resilience in a Turbulent World  high betrayal trauma    Betrayed: Why All Trauma Is Not Equal   March 18, 2013—When the topic of trauma comes up, we often wonder why some people are more resilient than others. \ldots & (psychological distress) AND (traumatic events) \\
17 & Protecting the professionals treating PTSD patients from burnout  How the Department of Veterans Affairs can take care of providers, so they can take care of patients.  For many mental healthcare clinicians at the US Department of Veterans Affairs (VA), listening to accounts of horrific, traumatic experiences is a routine part of the job. This type of work can be disturbing. There’s also another stressor VA mental health providers face: organizational politics and bureaucracy. Hector Garcia is \ldots & (vicarious trauma) AND (trauma exposure); (vicarious trauma) AND (trauma exposure) \\
18 & Journalist’s Guide to Coping withTraumatic Events  Turning Heads – Utilising Psychology  Covering conflicts can be too traumatic even for the seasoned journalist. What is scary is that many journalists often don’t know or are unaware of the psychological impact of their exposure to traumatic events. Some are unwilling to admit that they are affected by traumatic events that they witnessed in their coverage. The impact can be great on journalists and news crews that are on the forefront. \ldots & (psychological impact) AND (traumatic events) \\
19 & CHRIS 180 | CEU Georgia page-template,page-template-full\_width,page-template-full\_width-php,page,page-id-15632,page-child,parent-pageid-10,ajax\_fade,page\_not\_loaded,,qode-theme-ver-11.2,qode-theme-bridge,wpb-js-composer js-comp-ver-5.2.1,vc\_responsive  CHRIS 180 has transformed over 70,000 lives and has 36 years of experience providing highly individualized, trauma informed services that promote healing, recovery and resiliency. Our extensive experience is a solid demonstration of our \ldots & (vicarious trauma) AND (trauma exposure) \\
20 & Speaking About The Unspoken Existence Of PTSD Among Indigenous People  Anxiety, difficulty with sleep, sudden flashbacks coupled with heightened fear, feeling of being estranged, paranoia, and restlessness are dreading, if not almost fatal, symptoms of Posttraumatic Stress Disorder or PTSD. It is very common to hear about this mental health issue among people from the military.  Source: pixabay.com  Facts About PTSD  PTSD is a psychiatric disorder experienced by people who have been exposed to \ldots & (psychological distress) AND (trauma exposure) \\
\end{longtable}

\end{document}